\documentclass{article}
\usepackage{graphicx,latexsym}

\def\pd{\partial}
\def\a{\alpha}
\def\b{\beta}
\def\dl{\delta}
\def\s{\sigma}
\def\vphi{\varphi}

\def\lam{\lambda}
\def\Lam{\Lambda}

\def\Gm{\Gamma}
\def\om{\omega}

\def\hg{\hat{g}}
\def\bg{\bar{g}}

\def\bnabla{\bar{\nabla}}

\def\bR{\bar{R}}

\def\bE{\bar{E}}
\def\bC{\bar{C}}
\def\bG{\bar{G}}

\def\bDelta{\bar{\Delta}}
\def\P{{\rm P}}
\def\QG{{\rm QG}}
\def\D{{\rm D}}

\def\sq{\sqrt}
\def\e{\hbox{\large \it e}}
\def\half{\frac{1}{2}}
\def\fr{\frac}
\def\pp{\prime}

\def\bT{{\bf T}}

\def\bk{{\bf k}}

\def\bx{{\bf x}}

\def\lap3{~| \!\!\! \partial^2}
\def\dlap3{~| \!\!\! \partial^4}
\def\dH{{\dot H}}
\def\ddH{{\ddot H}}
\def\dddH{\stackrel{...}{H}}
\def\df{{\dot f}}
\def\ddf{{\ddot f}}
\def\dddf{\stackrel{...}{f}}
\def\dPsi{{\dot \Psi}}
\def\ddPsi{{\ddot \Psi}}
\def\dddPsi{\stackrel{...}{\Psi}}
\def\dPhi{{\dot \Phi}}
\def\ddPhi{{\ddot \Phi}}
\def\dddPhi{\stackrel{...}{\Phi}}
\def\ddddPhi{\stackrel{....}{\Phi}}
\def\dom{{\dot \omega}}
\def\ddom{{\ddot \omega}}
\def\dddom{\stackrel{...}{\omega}}
\def\ddddom{\stackrel{....}{\omega}}
\def\bb{\begin{equation}}
\def\ee{\end{equation}}
\def\bba{\begin{eqnarray}}
\def\eea{\end{eqnarray}}

\begin{document}

\begin{titlepage}

\begin{flushright}
{\sc KEK-TH-1098} \\
{\sc July 2006}
\end{flushright}

\begin{center}
{\Large {\bf Space-time Evolution and CMB Anisotropies \\ 
from Quantum Gravity}}
\end{center}

\vspace{5mm}

\begin{center}
{\sc Ken-ji Hamada\footnote{e-mail: hamada@post.kek.jp}, Shinichi Horata\footnote{e-mail: horata\_shinichi@soken.ac.jp} and Tetsuyuki Yukawa\footnote{e-mail: yukawa@soken.ac.jp}}
\end{center}

\begin{center}
{}$^1${\it Institute of Particle and Nuclear Studies, KEK, Tsukuba 305-0801, Japan} \\
{}$^1${\it Department of Particle and Nuclear Physics, The Graduate University for Advanced Studies (Sokendai), Tsukuba 305-0801, Japan} \\
{}$^{2,3}${\it Hayama Center for Advanced Research, The Graduate University for Advanced Studies (Sokendai), Hayama 240-0193, Japan}
\end{center}

\begin{abstract}
We propose an evolutional scenario of the universe which starts from quantum states with conformal invariance, passing through the inflationary era, and then makes transition to the conventional Einstein space-time. The space-time dynamics is derived from the renormalizable higher-derivative quantum gravity on the basis of a conformal gravity in four dimensions. Based on the linear perturbation theory in the inflationary background, we simulate evolutions of gravitational scalar, vector and tensor modes, and evaluate the spectra at the transition point located at the beginning of the big bang. The obtained spectra cover the range of the primordial spectra for explaining the anisotropies in the homogeneous CMB.
 
\vspace{5mm}

\end{abstract}
\end{titlepage}

\section{Introduction}
\setcounter{equation}{0}
\setcounter{figure}{0}
\noindent

After passing many theoretical as well as experimental tests, the general theory of relativity has been established as the fundamental theory of gravity capable of describing the universe. On the other hand, tracing the history of the universe, the space-time in its early epoch would be totally fluctuating quantum mechanically so that geometry lose its classical meaning. This would imply that there is a transition from quantum space-time to classical space-time, and there should exist a dynamical scale separating these two phases.

There is a possibility to observe the instance of the transition, because we can trace the past guided by known physical laws as far as the classical general relativity holds. Valuable information on physical processes taking place in the expanding universe has been recorded in the cosmic microwave background radiation (CMB) as tiny anisotropies. Angular power spectra of the anisotropies, recently observed by the Cosmic Background Explorer (COBE) \cite{cobe} and the Wilkinson Microwave Anisotropy Probe (WMAP) \cite{wmap,wmap3}, are, roughly speaking, projection of the history of universe for the period between its birth to the present. It gives us an amazing hope that if we believe the idea of inflationary universe \cite{guth, starobinsky} meaning an extremely rapid expansion without global thermalization, the long-distance correlation in observed anisotropies can provide information about dynamics of the period before the universe grew to the Planck scale. We are now at the stage of revealing and verifying the quantum aspect of universe.

The model of space-time transition proposed in this paper emerges from the renormalizable higher-derivative quantum theory of gravity developed on the basis of the conformal gravity in four dimensions. In this theory, at very high energies beyond the Planck scale, quantum fluctuations of the conformal mode in the metric field are dominated, and it is treated non-perturbatively. The space-time is described by a conformal field theory whose dynamics is governed by conformal invariant gravitational actions. The conformal field theory loses its validity at the dynamical scale indicated by the asymptotic freedom of the unique dimensionless coupling constant introduced for the traceless tensor mode. At about this point the universe is expected to make a transition from the quantum space-time to the classical Einstein space-time.

There are three mass scales in the model, namely the Planck mass $M_\P$, the dynamical scale $\Lam_\QG$, and the cosmological constant $\Lam_{\rm COS}$. We set their ordering as
\bb
      M_\P \gg \Lam_\QG \gg \Lam_{\rm COS}^{1/4}.
\ee
We shall obtain an evolutional scenario according to the order starting inflation driven by quantum effects of gravity without adding any artificial field by hands \cite{hy}. The inflationary model induced by quantum effects of gravity was first proposed by Starobinsky \cite{starobinsky}. We here develope along the idea and propose an evolutional scenario of the universe summarized as follows: the conformal symmetry begins to be broken about the Planck scale to form an inflationary universe with the expansion time constant of order of the Planck mass, and completely broken at the dynamical scale turning to the classical Einstein universe. Then, energies stored in extra degrees of freedom in higher-derivative gravitational fields shift to matter degrees of freedom, causing the big bang. The primordial power spectra obtained from the two-point correlation functions of gravitational fields show a significant character of the transition which is expected to be observed cosmologically.

The aim of this paper is to clarify why we can observe the Planck scale phenomena today from the tiny CMB anisotropies. Evolution of the universe is described by the equations of motion taking effects of running coupling as a time-dependent function. We find an inflationary solution that starts from the Planck scale and end at the dynamical scale where the transition of space-time occurs. Since the inflationary homogeneous solution is stable, the fluctuations from this solution will be slowly diminishing during the inflationary era, and thus the linear perturbation about the inflationary solution becomes applicable. We evaluate gravitational fluctuations perturbed about the homogeneous solution: the scalar perturbation so-called Bardeen potential, the vector perturbation and the tensor perturbation. We obtain the spectra at the transition point, which should be used for the primordial spectra to analize the data observed by WMAP. The rest of the paper organized as follows: we summarize the model of quantum gravity in section 2 and construct an evolutional scienario of the universe in section 3. In section 4 we study the linear perturbation theory in the inflationary background. We then simulate evolutions of perturbations and obtain primordial spectra in section 5. We conclude in section 6.

%%%%%%%%%%%%%%%%%%%%%%%%%%%%%%%%%%%%%%%%%%%%%%
%%%%   Renormalizable Quantum Gravity   %%%%
%%%%%%%%%%%%%%%%%%%%%%%%%%%%%%%%%%%%%%%%%%%%%%
\section{Renormalizable Quantum Gravity}
\setcounter{equation}{0}
\setcounter{figure}{0}
\noindent

Since it has been recognized that any attempt to quantize Einstein gravity perturbatively \cite{dewitt,fv,tv} cannot be succeeded \cite{gs}, most of researchers in this field feel the necessity of quantizing gravity in either a non-perturbative way or others. Historically, the renormalization problem is tackled introducing four-derivative terms to the Einstein-Hilbert action \cite{stelle, tomboulis, ft}, so that the gravitational coupling constant becomes dimensionless, and at the same time we can avoid the unboundness problem. Among various models, we employ the conformal gravity without $R^2$ term \cite{hamada02, nova}.

\paragraph{The Model} 
The conformal gravity is defned by the action,
\bba
   I &=& \int d^4 x \sq{-g} \biggl\{ 
         -\fr{1}{t^2} C^2_{\mu\nu\lam\s} -b G_4 + \fr{M_\P^2}{2}R -\Lam_{\rm COS} 
            \nonumber \\
     && \qquad\qquad\quad
         -\fr{1}{4} Tr \left( F^2_{\mu\nu} \right) 
         -\half \left( \pd^\mu X \pd_\mu X +\fr{1}{6}R X^2 \right) 
         + \cdots \biggr\},
            \label{I}
\eea  
where we write the reduced Planck mass and the cosmological constant as $M_\P =1/\sq{8\pi G}$ and $\Lam_{\rm COS}$, respectively. There are two conformal invariant four derivative actions in four dimensions: the square of the Weyl tensor $C^2_{\mu\nu\lam\s}$ and the Euler term $G_4$. The Weyl tensor is the field strength for the traceless tensor mode, and $t$ is the dimensionless coupling constant. The constant $b$ is introduced to renormalize divergences proportional to the Euler term, which is not an independent constant because it does not have a kinetic term.

The third term is the Einstein-Hilbert action, which will dominate at energies below the Planck scale. $F_{\mu\nu}$ is a gauge field strength, and $X$ is a scalar field with conformal coupling. In addition to these fields, we may consider fermions, which are conformally invariant in any dimensions, without mass terms.

The beta function of the renormalized coupling constant $t_r$ indicates asymptotic freedom. This implies that at high energies beyond the Planck scale, the coupling constant gets small, and accordingly configurations with the vanishing Weyl tensor will dominate. Then, the singular configulation such as a black hole at which the Riemann-Christoffel curvature tensor is divergent is excluded quantum mechanically. This is a prominent feature of the renormalizable quantum gravity based on the conformal gravity, in contrast to the theory based on the Einstein-Hilbert action because such a configuration has the vanishing scalar curvature so that its quantum weight in the path integral is unity.

The partition function is defined by the functional integral over the metric field $g_{\mu\nu}$. We here decompose the metric field into the conformal mode $\phi$, the traceless mode $h^\lam_{~\nu}$ and the background metric $\hg_{\mu\lam}$ as
\bb
    g_{\mu\nu}=\e^{2\phi}\bg_{\mu\nu}, 
\ee
and
\bb
    \bg_{\mu\nu}=\bigl( \hg \e^{h} \bigr)_{\mu\nu} 
    =\hg_{\mu\lam} \left( \dl^\lam_{~\nu} + h^\lam_{~\nu} +\cdots \right),
        \label{h-expansion}
\ee
with $tr(h)=h^\lam_{~\lam}=0$. Since we will consider a case with large running coupling constant, we include the coupling $t$ in the traceless mode $h^\lam_{~\nu}$ instead of $th^\lam_{~\nu}$. The signature of the metric is taken as $(-1,1,1,1)$. The contraction of the indices of $h_{\mu\nu}$ is done by using the background metric $\hg_{\mu\nu}$. In the following, quantities with the hat and the bar on them are defined in terms of the metric $\hg_{\mu\nu}$ and $\bg_{\mu\nu}$, and the contraction of them is done by using the metric $\hg_{\mu\nu}$ and $\bg_{\mu\nu}$, respectively. Unless it is not specified, the flat background $\hg_{\mu\nu}=\eta_{\mu\nu}$ is used, which defines the comoving frame with the coordinate $x^\mu=(\eta, x^i)$ and $\pd_\mu =(\pd_\eta, \pd_i)$.

Techniques to treat diffeomorphism invariance are developed in the two dimensional theory at the end of 1980's \cite{polyakov,kpz,dk,seiberg}, and extended later in four dimensions \cite{riegert, am, amm92, amm-phys, hamada99, hamada02,hh,hamada04,nova}. We change the path-integral measures from the diffeomorphism invariant measures to the practical measures defined on the background metric $\hg_{\mu\nu}$. In order to recover the diffeomorphism invariance the Wess-Zumino term $S$ \cite{wz} related to conformal anomalies \cite{cd, ddi, duff, bcr} in the action is necessary as the Jacobian, and the partition function is expressed as
\bb
     Z = \int \fr{[d\phi dh dA dX]_{\hg}}{\rm Vol(diff.)} 
          \exp \left\{ iS(\phi,\bg)+ iI(A,X,g) \right\}.
\ee
The induced action $S$ contains the kinetic term for the conformal field. At the lowest order in the coupling, it is given by the Riegert action \cite{riegert}
\bb
    S(\phi,\bg) = - \fr{b_1}{(4\pi)^2} \int d^4 x \sq{-\bg} \left\{
        2 \phi \bDelta_4 \phi +  \left( \bG_4 - \fr{2}{3} \bnabla^2 \bR \right) \phi 
        + \fr{1}{18} \bR^2 + \cdots \right\},
                  \label{Riegert}
\ee
where $\sq{-g} \Delta_4$ is a conformal invariant fourth order self-adjoint operator for a scalar field variable and $\nabla^2=\nabla^\lam \nabla_\lam$. The coefficient $b_1$ has been computed within the lowest order as \cite{amm92}
\bb
    b_1 = \fr{1}{360} \left( N_{\rm X} +\fr{11}{2} N_{\rm W} +62 N_{\rm A} \right) +\fr{769}{180},
               \label{b1}
\ee
where $N_{\rm X}$, $N_{\rm W}$ and $N_{\rm A}$ are the numbers of scalar fields, Weyl fermions and gauge fields added to the action, respectively. The dots indicate new vertices like $\phi^{n+1} \bDelta_4 \phi$, $\phi^n \bC_{\mu\nu\lam\s}^2$, and $\phi^n Tr(F_{\mu\nu}^2)$ induced at higher order of the coupling \cite{hamada02,hathrell}.

Since the conformal variation of the metric $\bg_{\mu\nu}$ is equivalent to the variation of the conformal field, the trace of derived stress tensor must be equal to the equation of motion of the confromal field. The second and third terms in the Riegert action (\ref{Riegert}) ensure such a diffeomorphism invariance condition within the linear approximation discussed in Section 4.

\paragraph{Running coupling constant}
The beta function of the renormalized coupling $t_r$ for the traceless mode has been calculated to be $\b =-\b_0 t^3_r$ with $\b_0 = \{ ( N_{\rm X} +3N_{\rm W} +12N_{\rm A} )/240 +197/60 \}/(4\pi)^2$ within the lowest order of perturbation \cite{ft,hamada02}. It indicates the dynamics of the traceless tensor mode asymptotically free. The effective action in the momentum space is expanded in $t_r$ as
\bba
      {\cal L}_{\rm eff} &=&  -\left\{ \fr{1}{t_r^2} - 2\b_0 \phi
         + \b_0 \log \left( \fr{k^2}{\mu^2} \right) +\cdots \right\} C^2_{\mu\nu\lam\s}
                  \nonumber \\
          &=& -\fr{1}{t_r^2(p)} C^2_{\mu\nu\lam\s}
\eea
for the Weyl tensor term, where $k$ is a comoving momentum defined on the flat background. The running coupling constant is approximately written as
\bb
  t^2_r(p) = \fr{1}{\b_0 \log (p^2/\Lam_\QG^2)} ,
      \label{running-mom}
\ee
where $p$ is a physical momentum defined by $p=k/\e^{\phi}$. The coupling constant is a measure of the degree of deviation from conformal field theory. The scale parameter $\Lam_\QG =\mu \exp \{ -1/2\b_0 t_r^2 \}$ with $\mu$ being a renormalization mass scale represents the energy scale where the metric field changes its appearance from the conformal mode and the traceless mode to the conventional Einstein gravity.

As discussed later, in the inflationary era the conformal field, $\phi$, grows large and the physical momentum diminishes exponentially. Accordingly, the running coupling constant $t_r(p)$ gets large during the inflation. As energies become lower to the dynamical scale $\Lam_\QG$, the running coupling constant diverges and the Weyl terms in the effective action will disappear. There, the conformal invariance terminates and the background metric appears as a physical measure of space and time. Beyond the energy smaller than this point the Einstein gravity dominates the dynamics.

%%%%%%%%%%%%%%%%%%%%%%%%%%%%%%%%%%%%%%%%%%%%%%%%%%%%%%
%%%%   Evolutional Scenario of The Universe  %%%%
%%%%%%%%%%%%%%%%%%%%%%%%%%%%%%%%%%%%%%%%%%%%%%%%%%%%%%
\section{Evolutional Scenario of The Universe}
\setcounter{equation}{0}
\setcounter{figure}{0}
\noindent

The renormalizable conformal quantum gravity suggests that there are four stages in the evolution of the universe divided by three mass scales. The first is quantum phase dominated by the dynamics of the conformal gravity action, which gradually turns to the second stage, i.e. the inflationary era. The third stage is the Einstein universe mostly ruled by the classical theory of general relativity. The last is the present deSitter evolution where the universe expanded sufficiently and the cosmological constant becomes effective. The aim of this section is to construct an evolutional model connecting the quantum gravity phase and the Einstein space-time.

\paragraph{Inflationary phase}
In the very early epoch when the space extends much smaller than the Planck scale, typical energy is higher than $M_\P$. Corrections of order $t_r^2$ in the asymptotic free regime can be neglected and the dynamics is governed by conformal field theory of gravity. As the universe expands typical energy gets lowered to $M_\P$, the Einstein action gradually becomes effective, so that the conformal symmetry starts to be broken to develop the classical solution. The effective action valid for such an energy regime is given by the sum of the Riegert action (\ref{Riegert}), the Einstein action and the matter action.  The homogeneous equations of motion has a stable inflationary solution \cite{hy}. At the $\Lam_\QG$ scale, this symmetry is completely broken and the universe make a transition to the classical space-time. In this section, we consider a dynamical effect arising from the effective action under the inflationary background.

The equations of motion simplified in the local form are obtaind in the way that the coefficient of the Riegert action of order $t_r^2$ corrections is written as \cite{nova}
\bb
    b_1 \rightarrow b_1 \left( 1 - a_1 t_r^2 + \cdots \right) 
    = b_1 B_0(t_r),
\ee
where $a_1$ is a positive constant, and $B_0$ is assumed to be 
\bb
          B_0(t_r) = \fr{1}{(1+\fr{a_1}{\kappa}t_r^2)^\kappa}. 
          \label{B0}
\ee
Here, $\kappa$ is a parameter for taking into account of the higher order perturbative effects phenomenologically, which lies in the range: $0 < \kappa \leq 1$. The equation of motion for the conformal mode is then obtained as
\bb
        - \fr{b_1}{4\pi^2} B_0 \pd_\eta^4 \phi   
                + M_{\rm P}^2\e^{2\phi} \left\{ 
              6 \pd_\eta^2 \phi                 +6 \pd_\eta \phi \pd_\eta \phi        
                           \right\}   =0 .  
                  \label{homogeneous1}
\ee
This equation comes from the trace part of the stress tensor. Here, there is no contribution from the Weyl and the matter terms, which are trivially traceless. The energy conservation equation from the time-time component of the stress tensor is given by
\bb
     \fr{b_1}{8\pi^2} B_0 \left\{
               2 \pd_\eta^3 \phi \pd_\eta \phi -\pd_\eta^2 \phi \pd_\eta^2 \phi   \right\} 
      - 3 M_{\rm P}^2 \e^{2\phi} \pd_\eta \phi \pd_\eta \phi  + \e^{4\phi} \rho 
      = 0.
              \label{homogeneous2}
\ee
Since the matter sector is conformal invariant, we here assume the stress tensor to be a form of the so-called perfect fluid with density $\rho$, although its contents may be different from that of ordinary matters. The contribution from the Weyl action also vanishes in the energy conservation equation.

It is known that these equations have a stable inflationary solution for the region where the coupling constant $t_r$ is small. As mentioned in section 2, however, the coupling constant can not stay small in the inflationary space-time. We must take into account the gradual increase of the coupling constant along with the expansion. The inclusion of the dynamical effects is not simple, and the equations of motion become rather complicated containing non-linear as well as non-local terms. Bravely, we simplify it under the following physical consideration: the running coupling constant $t_r(p)$ is an operator which acts on the field $\phi$, and it may fluctuate very much even on a smooth background. When the scale of local fluctuation of the order $1/M_\P$ is small compared to the size of the system, $1/\Lam_\QG$, we approximate the running coupling operator by its average under the spirit of the mean-field approximation. We rewrite the running coupling operator by time-dependent average replacing the physical momentum by $1/\tau$ in equation (\ref{running-mom}):
\bb
       t_r^2(\tau) =\fr{1}{\b_0 \log (1/\tau^2\Lam_\QG^2)},
          \label{running-time}
\ee
where $\tau$ is proper time defined by $d\tau =a(\eta)d\eta$ with $a=\e^\phi$. It shows that the running coupling diverges at the dynamical time scale, $\tau=\tau_\Lam$ with  $\tau_\Lam = 1/\Lam_\QG$.

Replacing the constant $t_r^2$ in the dynamical coefficient (\ref{B0}) by the time-dependent running coupling constant (\ref{running-time}), and introducing the variable $H =\dot{a}(\tau)/a(\tau)$, where the dot denotes the derivative with respect to the proper time $\tau$, the trace equations of motion (\ref{homogeneous1}) is written as
\bb
    \fr{b_1}{8\pi^2} B_0(\tau) \left( \dddH +7H\ddH +4\dH^2 +18H^2\dH +6H^4 \right)
     -3 M_\P^2 \left( \dH +2H^2 \right) =0 
        \label{vacuum-trace}
\ee
and the conservation equation (\ref{homogeneous2}) is
\bb
     \fr{b_1}{8\pi^2} B_0(\tau) \left( 2H\ddH -\dH^2 +6H^2\dH +3 H^4 \right) -3 M_\P^2 H^2
        + \rho =0.
\ee

For the region where the running coupling is small, these equations have a stable solution in which the scale factor exponentially grows up with the expansion time constant $H \simeq H_\D$, such that $a(\tau) \simeq \e^{H_\D \tau}$, where 
\bb
     H_\D=\sq{\fr{8\pi^2}{b_1}}M_\P. 
        \label{desitter}
\ee
As the coupling constant gradually increases, the value of the function $B_0$ reduces, and the value of $H$ and its time derivatives increase. At the transition point $\tau_\Lam$ for $0 < \kappa \leq 1$, the the third derivative of $H$ diverges. In the case of $\kappa =1$, the second derivative of $H$ also diverges. The combination $B_0 \ddH$, however, vanishes in any case at the transition point so that the energy density is kept finite.

The energy conservation equation implies that the density $\rho$ starts almost vanishing in the inflationary era, and at the transition point it sharply increases to the finite value $\rho(\tau_\Lam) = 3M_\P^2 H^2(\tau_\Lam)$.\footnote{%%%%%%%%%%%(footnote)%%%%%%%%
This value is consistent with the radiation density for the Hawking temperature of deSitter space-time propotional to $H_\D$, although the meaning of the temperature in quantum space-time seems obscure.
} %%%%%%%%%%%%%% 
This behavior becomes more clear when we consider the time derivative of matter density,
\bb
    \dot{\rho} +4 H\rho = \fr{b_1}{8\pi^2} \dot{B}_0(\tau) 
               \left( 2H\ddH -\dH^2 +6H^2 \dH +3H^4 \right) . 
\ee
Initially $\dot{B}_0 \simeq 0$, and the matter density is not increasing, while it increases sharply about the transition point where $B_0$ changes drastically.

\begin{figure}[t]
\begin{center}
\includegraphics{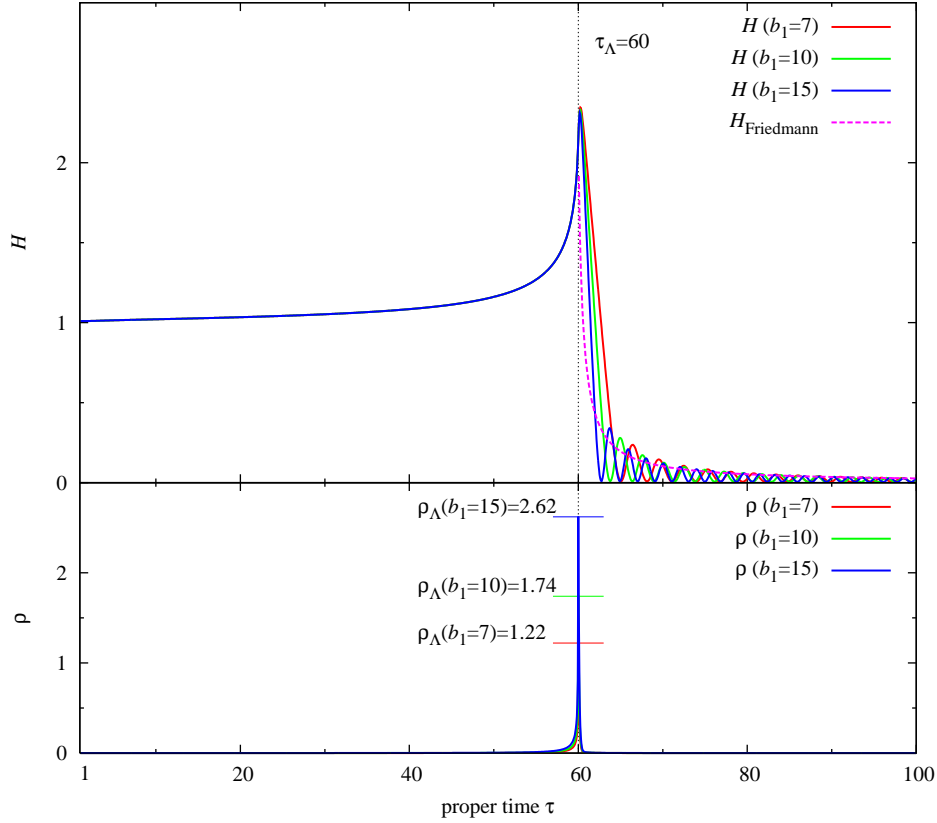}
\end{center}
\caption{\label{fig 3.1} The time evolution of $H$ and $\rho$ for $b_1=7$, $10$, and $15$. $H_\D$ is normalized to be unity, and thus $\tau_\Lam=60$ and inflationary solution for $H$ becomes independent of $b_1$.  The dashed line denotes the Friedmann solution for $H$.}
\end{figure}

\begin{figure}[t]
\begin{center}
\includegraphics{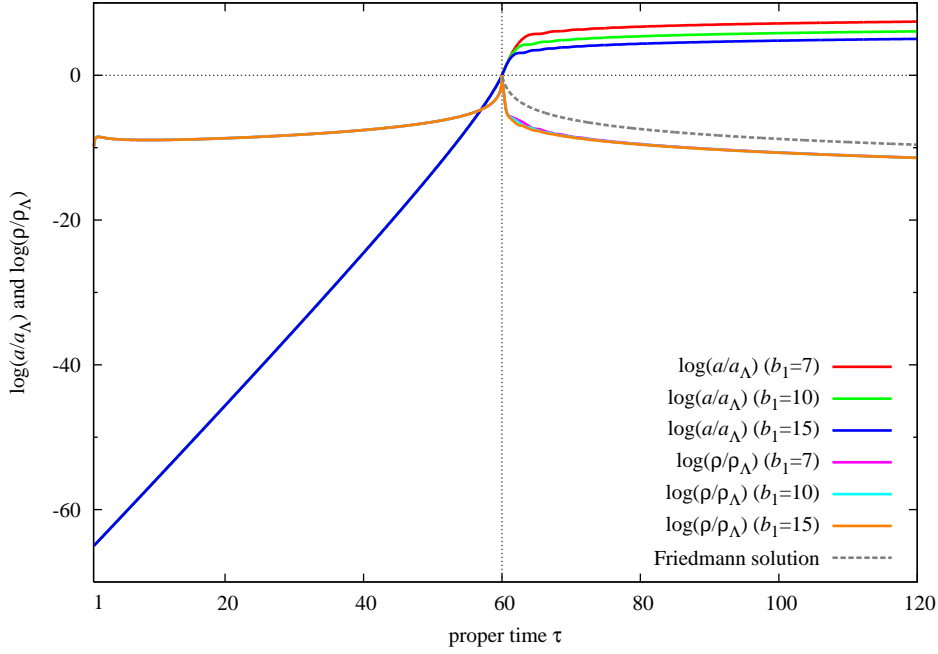}
\end{center}
\caption{\label{fig 3.2} The time evolution of $a$ and $\rho$ for $b_1=7$, $10$, and $15$ in a logarithmic plot, normalized by their values at $\tau_\Lam~(=60)$ denoted by $a_\Lam$ and $\rho_\Lam$, respectively. The dashed line denotes the Friedmann solution for $\rho$.}
\end{figure}

Solving equations of the motion numerically, we obtain the solutions for $H$ and $\rho$ shown in figure \ref{fig 3.1}, and $a$ and $\rho$ in figure \ref{fig 3.2}. The simulation is performed starting at the Planck time defined by $\tau_\P=1/H_\D$ and ending at $\tau_\Lam=1/\Lam_\QG$. We choose three values $b_1=7,~10,~15$ for the coefficient of the Riegert action determined from the conformal field, where the values correspond to the Standard model and various GUT models.\footnote{ %%%%%%%(footnote)%%%%
For example, $b_1=7.0~(N_{\rm A}=12,~N_{\rm W}=45)$ for the Standard model,  $b_1=9.1$ for $SU(5)~(N_{\rm A}=24, N_{\rm W}=45)$, $b_1=12.0$ for $SO(10)~(N_{\rm A}=45)$, and $b_1=17.7$ for $E_6~(N_{\rm A}=78)$.  
}  %%%%%%%%%%%%
The other parameters are rather indistinct because they depend on the non-perturbative dynamics of the traceless mode, and they are likely to be determined phenomenologically. Here, they are chosen as $\beta_0=0.6$, $a_1=0.1$ and $\kappa=0.5$. The ratio of the mass scales is set as $H_\D/\Lam_\QG =60$ and by this choice the number of e-foldings from the Planck time to the dynamical time becomes 
\bb
    {\cal N}_e=\log \fr{a(\tau_\Lam)}{a(\tau_\P)}=65.
       \label{e-foldings}
\ee
This e-foldings gives the desirable number large enough to explain the evolutional scenario of the universe from the Planck time to the present and the primordial spectrum discussed in section 5.

This simulation suggests that the quantum fluctuation of the conformal field covering over the whole quantum universe makes a transition to matter fluctuations at $\tau_\Lam$, and the expansion slows down from the inflating conformal space-time to the Einstein space-time as discussed below. The entropy is then generated to develop the thermal universe, which is regarded as the big bang in our scenario. The primordial spectrum we extract from the CMB observation is expected to be the reflection of quantum fluctuation of the gravitational field developed right before the transition.

%%%%%%%%%%%%%(Figure 3.1 input)%%%%%
%%%%%%%%%%%%%%%%%%%%%%%%%%%%%%%%%

%%%%%%%%%%%%%(Figure 3.2 input)%%%%%
%%%%%%%%%%%%%%%%%%%%%%%%%%%%%%%%%

\paragraph{Einstein phase}
Below the energy scale $\Lam_\QG$, the Einstein action becomes dominated, and the space-time makes transition to the Friedmann universe. Here, in analogy to the chiral perturbation theory as the low energy effective theory of QCD \cite{dgh}, we derive the low energy effective theory of gravity valid below the energy scale $\Lam_\QG$, and connect the solution in the inflationary phase with that in the Einstein phase smoothly at the transition point.

In the case of QCD, dynamics of gauge fields disappears below the dynamical QCD scale, and meson and baryon become dynamical fields. In quantum gravity, although dynamics of conformal gravity disappear below $\Lam_\QG$, the metric tensor still remains as the dynamical variable in the Einstein gravity. As a consequence, the low energy effective action is given by an expansion in derivatives of the metric field as
\bb
     I_{\rm low} = \int d^4 x \sq{-g}\left\{ {\cal L}_2 + {\cal L}_4  + \cdots \right\},
\ee
where the lowest derivative cosmological term is not included because it is negligible at the energy scale we are discussing. The two derivative term is the Einstein action 
\bb
     {\cal L}_2 = \fr{M_\P^2}{2} R + {\cal L}_2^{\rm M},
\ee
where ${\cal L}_2^{\rm M}$ is the conformally invariant matter lagrangian.

In practice we employ terms up to the fourth order. Following the prescription of the low energy effective theory, ${\cal L}_2$ is used for calculating both tree and one loop diagrams, while ${\cal L}_4$ is used for the tree diagrams. The Planck mass is then regarded as the counterpart of the pion decay constant in chiral perturbation theory. The higher derivative terms are expanded by the inverse of the Planck constant, which can be justified owing to the ordering $M_\P \gg \Lam_\QG$. The possible terms for ${\cal L}_4$ are given by
\bb
       R^2, \quad  R^2_{\mu\nu}, \quad R_{\mu\nu\lam\s}^2,  \quad
       \fr{1}{M_\P^2} R_{\mu\nu} T_{\rm M}^{\mu\nu}, \quad 
       \fr{1}{M_\P^4}T_{\rm M}^{\mu\nu} T^{\rm M}_{\mu\nu}, 
\ee
where $T^{\rm M}_{\mu\nu}$ is the stress tensor for conformally invariant matters.

Since we restrict our effective action up to the fourth order, the Einstein equations of motion $M^2_\P R_{\mu\nu}=T^{\rm M}_{\mu\nu}$ is used to reduce the terms in ${\cal L}_4$. The equation of motion also implies $R=0$. Taking into account these equations and Euler relation to remove the square of the Riemann-Christoffel curvature tensor, the independent terms in ${\cal L}_4$ is simply given by
\bb
       {\cal L}_4 = \fr{\a}{(4\pi)^2} R^{\mu\nu} R_{\mu\nu} .
\ee
Here, the constant $\a$ is a positive value which is phenomenologically determined.

The constant $\a$ acquires loop corrections coming from ${\cal L}_2$. The renormalization procedure is done by background field method \cite{dewitt,tv} about the background satisfying the Einstein equations of motion with the cutoff $\Lam~(<\Lam_\QG)$. The constant $\a$ receives a finite renormalization dependent on the cutoff scale.  The differences between two cutoff scales, $\Lam$ and $\Lam_{\QG}$, will be given by 
\bb
    \a(\Lam) =\a(\Lam_\QG)+\zeta \log(\Lam^2/\Lam_\QG^2) .
    \label{coupling-alpha}
\ee 
Here, $\zeta$ is a gauge invariant constant, because the Ricci tensor is divergence-free by equation of motion: $\nabla^\mu R_{\mu\nu} = \nabla^\mu T^{\rm M}_{\mu\nu}=0$. For the contributions from diagrams with internal matter loops, $\zeta = (N_{\rm X} + 3 N_{\rm W} +12 N_{\rm A})/120$. Equation (\ref{coupling-alpha}) shows that the coupling $\a$ becomes small and the four-derivative term becomes irrelevant at low energies. It is worth commenting that this theory is valid at the low energy below $\Lam_\QG$, and therefore a ghost pole located at the order of $M_\P$ does not emerge. Therefore, the higher derivative action in the low energy effective action does not conflict with unitarity.

We firstly solve the equation of motion for the homogeneous component in the low energy effective theory. It is given by the trace of the stress tensor, or the variation of the conformal mode,
\bb
       M_\P^2 \left( \dH +2H^2 \right) 
       + \fr{\a}{4\pi^2} \left( \dddH +7H\ddH +4\dH^2 +12H^2\dH  \right) =0
          \label{trace-ein}
\ee
for $H$ as a function of the proper time. The energy conservation equation is obtained as
\bb
   -3 M_\P^2 H^2 + \rho + \fr{\a}{4\pi^2} \left( -6H\ddH +3\dH^2 -18H^2\dH \right) =0.
          \label{energy-ein}
\ee

When we restrict terms in the effective action up to the fourth order in derivatives, it would be valid at the energy scale sufficiently below $\Lam_\QG$. In order to fill the gap between $\Lam_\QG$ and the low energy scale $\Lam$, we naively interpolate the equation up to $\Lam_\QG$ by assuming the coupling to be time-dependent:
\bb
    \a(\tau) = \a_0 + \zeta \log \left( \fr{1}{\tau^2 \Lam_\QG^2} \right) 
       \simeq \fr{\a_0}{1+\fr{\zeta}{\a_0} \log(\tau^2\Lam^2_\QG)}, 
\ee
where $\a_0=\a(\Lam_\QG)$. This form is given by replacing the cutoff with the inverse of the proper time in (\ref{coupling-alpha}). By the running coupling we expect to take into account the space expansion effect as we have done for the   coupling $t_r(p) \rightarrow t_r(1/\tau)$ in the last section. By this coupling, it is expected to vanish at low energy, and it is written in an non-negative form.

Equations for both sides of the transition time $\tau_\Lam$ are the same order, and we can smoothly connect the Einstein space-time with the inflationary space-time. The initial values of $H$, $\dH$ and $\rho$ are chosen to connect with those in the inflationary phase. The initial value of $\ddH$ in order to solve equation ({\ref{trace-ein}) is given by solving the energy conservation equation (\ref{energy-ein}). We can show that the Friedmann universe, $\dH+2H^2=0$ and $3M_\P^2H^2=\rho$, is a solution of equations of motion (\ref{trace-ein}) and (\ref{energy-ein}), and general solutions are approaching to the Friedmann universe oscilating about this solution. The numerical simulation is carried out by setting the parameters as $\a_0=1$ and $\zeta=1$. The results are given in figures \ref{fig 3.1} and \ref{fig 3.2}, in which $H_\D$ is normalized to be unity such that the solutions depend on the value $b_1$. For a few steps in the period after the transition, the space-time expands acceleratingly and the matter density decreases sharply during that period. It soon reduces to the Friedmann universe.

%%%%%%%%%%%%%%%%%%%%%%%%%%%%%%%%%%%%%%%%
%%%%   Linear Perturbationn Theory  %%%%
%%%%%%%%%%%%%%%%%%%%%%%%%%%%%%%%%%%%%%%%
\section{Linear Perturbation Theory}
\setcounter{equation}{0}
\setcounter{figure}{0}
\noindent

Since the inflation is a stable solution, it is expected that the amplitude of fluctuation of scalar curvature, $\dl R$, damps during the inflation, but does not vansish. The amplitude is estimated as follows: since scalar curvature has two derivatives, the size of the fluctuation at the Planck scale is estimated to be $\dl R \sim H_\D^2$ such that $\dl R/R \sim H_\D^2/12H_\D^2 \sim 10^{-1}$ in the inflationary background, where the denominator is normalized by the deSitter curvature with $H=H_\D$. Near the transition point, the running coupling gets large and the energy scale is the order of $\Lam_\QG$. Then, the the scalar curvature contrast at the transition point is estimated to be $ \dl R/R \sim \Lam_\QG^2/12H_\D^2$. If we take the dynamical mass smaller than the Planck mass by two digits, the amplitude becomes the observed order of $10^{-5}$. This allows linear perturbations \cite{bardeen,ks,hs,ll} about the inflationary solution applicable from the Planck scale to the dynamical transition scale.

The asymptotic freedom implies that the fluctuations of vector and tensor modes are relatively small compared to the scalar mode. In addition, the fluctuation we will discuss expands rapidly enough during the inflation to the size far from the horizon scale, and thus it is not affected by the dynamics near the transition point. The linear perturbation can be used for such a size of the vector and the tensor as well.

\paragraph{Perturbations}
The conformal mode and matter density are perturbed about the homogeneous solution as
\begin{eqnarray}
         \phi(\eta,{\bf x}) &=& \phi (\eta) + \vphi (\eta, {\bf x}), 
                  \nonumber \\
         \rho(\eta,{\bf x}) &=& \rho (\eta) +\dl \rho(\eta, {\bf x}),
                \label{non-homo-field}
\end{eqnarray}
where $\phi (\eta)$ and $\rho(\eta)$ are solutions of the equations of motion (\ref{homogeneous1}) and (\ref{homogeneous2}). In below, except in the appendix A, $\phi$ denotes the homogeneous solution.

The stress tensor for the matter sector is traceless, which is taken to be 
\bba
      \bT^{{\rm M}\lam}_{~~~\lam} &=& 0, 
              \nonumber \\
      \bT^{\rm M}_{00} &=& \e^{4\phi} (\rho +\dl \rho +4\rho \vphi), 
              \nonumber \\
      \bT^{\rm M}_{0i} &=& - \fr{4}{3} \e^{4\phi} \rho \left(v_i + \half h_{0i} \right), 
              \nonumber \\
      \bT^{\rm M}_{ij} &=& \e^{4\phi} \left\{ \fr{1}{3} (\rho +\dl \rho + 4\rho \vphi) \dl_{ij}
                                                        + \Pi_{ij} \right\},
             \label{T-matter}
\eea
where $v_i =\pd_i v +v^{\rm T}_i$ are velocity perturbations, where $v_i^{\rm T}$ is the transverse component. The anisotropic stress is traceless, $\Pi^i_{~i}=0$. The appearance of the conformal factor is due to the convention of the stress tensor defined in the appendix A.

Hence, we consider the linear perturbations of field variables, $\vphi(\eta, {\bf x})$, $h_{\mu\nu}(\eta,{\bf x})$, $\dl \rho(\eta,{\bf x})$, and $v_i(\eta,{\bf x})$.

\paragraph{Gauge invariance}
Under the general coordinate transformations, $\dl_\xi g_{\mu\nu}=g_{\mu\lam}\nabla_\nu \xi^\lam + g_{\nu\lam}\nabla_\mu \xi^\lam$, field variables are transformed within a linear approximation as
\bba
      \dl_\xi \vphi &=& \xi^\lam \pd_\lam \phi +\fr{1}{4} \pd_\lam \xi^\lam, 
              \nonumber \\
      \dl_\xi h_{\mu\nu} &=& \pd_\mu \xi_\nu + \pd_\nu \xi_\mu - \half \eta_{\mu\nu} \pd_\lam \xi^\lam,
               \label{gct}
\eea
where $\xi_\mu=\eta_{\mu\nu}\xi^\nu$. The traceless tensor mode is further decomposed as
\bba
  h_{00}&=&h, 
        \nonumber \\
  h_{0i}&=&h_i^{\rm T} +\pd_i h^\pp, 
        \nonumber \\
  h_{ij}&=&h_{ij}^{\rm TT}+\pd_{(i} h_{j)}^{{\rm T} \pp}  + \fr{1}{3}\dl_{ij}h
           + \left( \fr{\pd_i \pd_j}{\lap3} -\fr{1}{3} \dl_{ij} \right) h^{\pp\pp},     
\eea
where $i,j=1,2,3$. $h_i^{\rm T}$ and $h_i^{{\rm T}\pp}$ are the transverse vectors and $h_{ij}^{\rm TT}$ is the transverse-traceless tensor. $\lap3=\pd^i \pd_i$ is the spacial comoving Laplacian. If $\xi^\mu$ is decomposed as $\xi^0$ and $\xi_i = \xi_i^{\rm T} +\pd_i \xi^{\rm S}$, the general coordinate transformations are described in terms of $\xi$'s as
\bba
      \dl_\xi \vphi &=& \xi^0 \pd_\eta \phi + \fr{1}{4} \pd_\eta \xi^0 +\fr{1}{4} \lap3 \xi^{\rm S}, 
                     \nonumber \\
      \dl_\xi h &=& -\fr{3}{2} \pd_\eta \xi^0 + \half \lap3 \xi^{\rm S}, 
                     \nonumber \\
      \dl_\xi h^\pp &=& -\xi^0 + \pd_\eta \xi^{\rm S},  
                     \nonumber \\
      \dl_\xi h^{\pp\pp} &=& 2 \lap3 \xi^{\rm S}, 
                     \nonumber \\
      \dl_\xi h^{\rm T}_i &=& \pd_\eta \xi^{\rm T}_i, 
                     \nonumber \\
      \dl_\xi h^{{\rm T} \pp}_i &=& 2 \xi^{\rm T}_i, 
                     \nonumber \\
      \dl_\xi h^{\rm TT}_{ij} &=& 0.  
\eea
For the matter sector, the perturbed variables are transformed as
\bba
     \dl_\xi v &=& -\pd_\eta \xi^{\rm S}, 
              \nonumber \\
     \dl_\xi v_i^{\rm T} &=& -\pd_\eta \xi^{\rm T}_i, 
              \nonumber \\
     \dl_\xi (\dl \rho) &=& \xi^0 \pd_\eta \rho,
              \nonumber \\
     \dl_\xi \Pi^i_{~j} &=& 0.         
\eea

The gauge invariant gravitational potentials so-called Bardeen potentials are defined by 
\bba
   \Phi &=& \vphi + \fr{1}{6} h - \fr{1}{6} h^{\pp\pp} + \s \pd_\eta \phi , 
                  \nonumber \\
   \Psi &=& \vphi - \half h + \s \pd_\eta \phi  + \pd_\eta \s ,   
\eea
where
\bb
  \s = h^\pp -\half \fr{\pd_\eta h^{\pp\pp}}{\lap3}.
        \label{sigma-field}
\ee
If we take the gauge $h^\pp = h^{\pp\pp} =0$, the Bardeen potentials are expressed as $\Phi=\vphi+h/6$ and $\Psi=\vphi-h/2$ such that the metric has the following form:
\bb
  ds^2 = a^2 \left[ - \left( 1+ 2\Psi \right) d\eta^2 
                          + \left( 1+ 2\Phi \right) d{\bf x}^2 \right]
\ee
for the scalar perturbations. The gauge invariant vector and tensor perturbations are defined by
\bb
     \Upsilon_i = h_i^{\rm T} -\half \pd_\eta h_i^{{\rm T}\pp} 
     \quad \hbox{and}   \quad h_{ij}^{\rm TT}.
\ee

For the matter sector, gauge invariant perturbations are defined by
\bba
    D &=& \fr{\dl \rho}{\rho} + \fr{\pd_\eta \rho}{\rho} \s -4 \pd_\eta \phi V, 
                 \nonumber \\
    V &=&  v + \half \fr{\pd_\eta h^{\pp\pp}}{\lap3} ,
                 \nonumber \\
    V_i &=& v_i^{\rm T} + \half \pd_\eta h_i^{{\rm T}\pp} ,
                 \nonumber \\
    \Omega_i &=& v_i^{\rm T} + h_i^{\rm T}.
\eea
Here, the vector variables satisfy the relation $\Upsilon_i+V_i-\Omega_i=0$.

The gauge invariance restricts the form of the stress tensor with the dynamical factor for the conformal-field sector. In order to obtain the gauge invariant equations of motion, it is necessary to rewrite the function $B_0$ by the field-dependent function $B$ which transforms as a scalar under the coordinate transformation:
\bb
     \dl_\xi B  = \xi^\lam \pd_\lam B = \xi^0 \pd_\eta B. 
\ee
Using the $\s$ field, which is transformed as $\dl_\xi \s = -\xi^0$, the scalar function $B$ is written as
\bb
       B = B_0 - \s \pd_\eta B_0.
         \label{B}
\ee
The stress tensors for the Riegert action are given in the appendix A. Replacing the coefficient $b_1$ in the expressions by the dynamical factor $b_1 B$, we obtain the equations of motion for the conformal-field sector.

\paragraph{Linear scalar equations}
Let us write down the linear equations of motion for the gauge invariant scalar perturbations, in which the anisotropic stress tensor, $\Pi_{ij}$, does not contribute. We find four independent dynamical equations of motion for four independent scalar perturbations, $\Phi$, $\Psi$, $D$, and $V$. We first consider the following two equations:
\bba
    \bT^\lam_{~\lam} &=& 0, 
         \label{equation1} \\
    \fr{1}{\lap3} \left( \bT^i_{~i}-3\fr{\pd^i\pd^j}{\lap3} \bT_{ij} \right) &=& 0.  
         \label{equation2}
\eea
These equations are independent of the matter sector, and we can obtain the Bardeen potentials by solving these equations. In terms of the gauge invariant variables, the trace equation (\ref{equation1}) can be written as
\bba
  && \fr{b_1}{8\pi^2} B_0(\tau) \biggl\{ 
         -2 \pd_\eta^4 \Phi                   -2 \pd_\eta \phi \pd_\eta^3 \Phi
         + \left( -8 \pd_\eta^2 \phi + \fr{10}{3} \lap3 \right) \pd_\eta^2 \Phi 
             \nonumber \\ 
  && \qquad\qquad
         + \left( -12 \pd_\eta^3 \phi +\fr{10}{3} \pd_\eta \phi \lap3 \right) \pd_\eta \Phi  
         + \left( \fr{16}{3} \pd_\eta^2 \phi  - \fr{4}{3} \lap3 \right) \lap3 \Phi
             \nonumber \\ 
  && \qquad\qquad 
         + 2 \pd_\eta \phi \pd_\eta^3 \Psi 
         + \left(8 \pd_\eta^2 \phi + \fr{2}{3} \lap3 \right) \pd_\eta^2 \Psi           
         + \left( 12 \pd_\eta^3 \phi - \fr{10}{3} \pd_\eta \phi \lap3 \right) \pd_\eta \Psi
             \nonumber \\ 
  && \qquad\qquad
         + \left( - \fr{16}{3} \pd_\eta^2 \phi  - \fr{2}{3} \lap3 \right) \lap3 \Psi
    \biggr\}
             \nonumber \\ 
  && 
    + M_\P^2 \e^{2\phi} \Bigl\{ 
           6\pd_\eta^2 \Phi                 + 18 \pd_\eta \phi \pd_\eta \Phi 
           - 4 \lap3 \Phi                   -6 \pd_\eta \phi \pd_\eta \Psi 
             \nonumber \\ 
  && \qquad\qquad\qquad    
         + \left( 12 \pd_\eta^2 \phi + 12 \pd_\eta \phi \pd_\eta \phi - 2\lap3 \right) \Psi 
           \Bigr\}  = 0 . 
           \label{equation-trace}  
\eea
Here, the field $\pd_\eta^4 \phi$ and the factor $\pd_\eta B_0$ in $B$ are removed by using the homogeneous equations of motion. The formulas in the appendix B are useful when equations of motion are transformed in terms of the proper time. The space-space component equation (\ref{equation2}) gives
\bba
   && \fr{b_1}{8\pi^2} B_0 (\tau) \biggl\{
          \fr{4}{3} \pd_\eta^2 \Phi           + 4 \pd_\eta \phi \pd_\eta \Phi
          + \left( \fr{28}{3} \pd_\eta^2 \phi -\fr{8}{3} \pd_\eta \phi \pd_\eta \phi 
                    -\fr{8}{9} \lap3 \right) \Phi
             \nonumber \\ 
   && \qquad\qquad\qquad
         - \fr{4}{3} \pd_\eta \phi \pd_\eta \Psi 
         + \left( -\fr{4}{3} \pd_\eta^2 \phi + \fr{8}{3} \pd_\eta \phi \pd_\eta \phi 
                   - \fr{4}{9} \lap3  \right) \Psi
     \biggr\}
            \nonumber \\
   && + \fr{2}{t_r^2(\tau)} \left\{ 4 \pd_\eta^2 \Phi -\fr{4}{3} \lap3 \Phi 
                              -4 \pd_\eta^2 \Psi + \fr{4}{3} \lap3 \Psi \right\}
             \nonumber \\ 
  && + M_\P^2 \e^{2\phi} \left\{
          - 2 \Phi          - 2 \Psi  
           \right\}  =0.
           \label{equation-space}
\eea
Equation (\ref{equation-space}) is of a second order with respect to the time derivative. This equation plays an important role in connecting between the inflation and the Einstein phases. In the limit $t_r \rightarrow 0$, where the conformal field dominates, $\Phi=\Psi$ is realized due to the vanishing of the Weyl tensor, while at the transition point where the coupling diverges, the configulation with $\Phi = -\Psi$ should be realized.

The matter density fluctuation $D$ is determined by the equation for $(00)$-component of stress tensor, and the velocity perturbation $V$ are determined by the equation for $(0i)$-component. We here consider the following two combinations:
\bba
   \bT_{00} + 3 \pd_\eta \phi \fr{\pd^i}{\lap3} \bT_{i0} &=& 0,  
          \label{equation3}  \\
   \fr{\pd^i}{\lap3} \bT_{i0} &=& 0. 
          \label{equation4}
\eea 
In terms of the gauge invariant variables, the energy conservation equation (\ref{equation3}) is written as
\bba
  && \fr{b_1}{8\pi^2} B_0(\tau) \biggl\{
          \left( -2 \pd_\eta^2 \phi      + 2 \pd_\eta \phi \pd_\eta \phi 
                 -\fr{2}{3} \lap3     \right) \pd_\eta^2 \Phi
          +\left( 2\pd_\eta^3 \phi   -4 \pd_\eta^2 \phi \pd_\eta \phi
                   \right) \pd_\eta \Phi
             \nonumber \\
  && \qquad\qquad
         + \pd_\eta \phi \left( -2 \pd_\eta^2 \phi    + 2 \pd_\eta \phi \pd_\eta \phi 
                                -2 \lap3  \right) \pd_\eta \Phi 
         + \left( -\fr{20}{3} \pd_\eta \phi \pd_\eta \phi  + \fr{4}{9} \lap3 \right) \lap3 \Phi 
        \nonumber \\
  && \qquad\qquad
         + \pd_\eta \phi \left( 2 \pd_\eta^2 \phi     - 2 \pd_\eta \phi \pd_\eta \phi 
                                + \fr{2}{3} \lap3 \right) \pd_\eta \Psi 
         + \left( -2 \pd_\eta^3 \phi \pd_\eta \phi    + 4 \pd_\eta^2 \phi \pd_\eta^2 \phi \right) \Psi
       \nonumber \\
  && \qquad\qquad
         + \left( 2 \pd_\eta^2 \phi     + \fr{2}{3} \pd_\eta \phi \pd_\eta \phi  
                  + \fr{2}{9} \lap3  \right) \lap3 \Psi 
        \biggr\}
             \nonumber \\
  && + \fr{2}{t_r^2(\tau)} \left\{ 
         - \fr{4}{3} \dlap3 \Phi                -4 \pd_\eta \phi \lap3 \pd_\eta \Phi
         + \fr{4}{3} \dlap3 \Psi                + 4 \pd_\eta \phi \lap3 \pd_\eta \Psi 
       \right\} 
             \nonumber \\ 
  && + M_\P^2 \e^{2\phi} 2 \lap3 \Phi   +\e^{4\phi} \rho D = 0.
           \label{equation-energy}
\eea
and equation (\ref{equation4}) is 
\bba
   && \fr{b_1}{8\pi^2} B_0(\tau) \biggl\{
             -\fr{2}{3} \pd_\eta^3 \Phi         
             + \left( -\fr{10}{3} \pd_\eta^2 \phi   +\fr{2}{3} \pd_\eta \phi \pd_\eta \phi    
                      + \fr{4}{9} \lap3 \right) \pd_\eta \Phi
             -\fr{4}{3} \pd_\eta \phi \lap3 \Phi 
               \nonumber \\
   && \qquad
             + \fr{2}{3} \pd_\eta \phi \pd_\eta^2 \Psi 
             + \left( 2 \pd_\eta^2 \phi           -\fr{2}{3} \pd_\eta \phi \pd_\eta \phi 
                      + \fr{2}{9} \lap3  \right) \pd_\eta \Psi
             + \left( 2 \pd_\eta^3 \phi       - \fr{2}{3} \pd_\eta \phi \lap3 \right) \Psi
        \biggr\} 
             \nonumber \\
   && + \fr{2}{t_r^2(\tau)} \left\{ 
            - \fr{4}{3} \lap3 \pd_\eta \Phi       + \fr{4}{3} \lap3 \pd_\eta \Psi  \right\}
               \nonumber \\
   && + M_\P^2 \e^{2\phi} \left\{ 
             2 \pd_\eta \Phi  -2 \pd_\eta \phi \Psi \right\} 
      - \fr{4}{3} \e^{4\phi} \rho V  =0.
        \label{equation-velocity}
\eea
These equations involve up to the third order in the time derivative. Thus, substituting the Bardeen potentials obtained from (\ref{equation-trace}) and (\ref{equation-space}) into these equations, we can get the density perturbation $D$ and velocity perturbations $V$. Since terms from the Riegert and the Weyl sectors with dynamical factors disappear at the transition point $\tau_\Lam$ according to the vanishing of the factors, $1/t_r^2$ and $B_0$, the values of $D(\tau_\Lam)$ and $V(\tau_\Lam)$ are determined from the value of $\Phi(\tau_\Lam)~(=-\Psi(\tau_\Lam))$. Thus, it is enough to compute the Bardeen potentials to achieve information necessary at the transition time $\tau_\Lam$.

\paragraph{Linear vector equations}
The vector fluctuation $\Upsilon_i$ is determined by the equation for $(ij)$-component of stress tensor, and the velocity perturbation $\Omega_i$ is determined by the equation for $(0i)$-component of stress tensor. We consider the following combinations:
\bba
        \fr{\pd^j}{\lap3}\bT_{ij} &=& 0,
            \label{equation5} \\
        \bT_{0i} &=& 0.
            \label{equation6}
\eea
Equation (\ref{equation5}) is written as 
\bba
  && \fr{2}{t_r^2(\tau)} \left\{  \pd_\eta^3 \Upsilon_i  -\pd_\eta \lap3 \Upsilon_i  \right\}
            \nonumber \\
  && -\fr{b_1}{8\pi^2}B_0(\tau) \left\{  
              \left( \fr{1}{3}\pd_\eta^2 \phi +\fr{4}{3} \pd_\eta \phi \pd_\eta \phi 
                         \right) \pd_\eta \Upsilon_i 
              +\left( \fr{1}{3} \pd_\eta^3 \phi + \fr{8}{3} \pd_\eta^2 \phi \pd_\eta \phi 
                         \right) \Upsilon_i 
         \right\}
            \nonumber \\
  && +M_\P^2 \e^{2\phi} \left\{ \half \pd_\eta \Upsilon_i + \pd_\eta \phi \Upsilon_i   \right\} =0.
           \label{equation-vector1}
\eea
and equation (\ref{equation6}) is
\bba
   && \fr{2}{t_r^2(\tau)} \left\{  \pd_\eta^2 \lap3 \Upsilon_i - \dlap3 \Upsilon_i \right\}
     -\fr{b_1}{8\pi^2}B_0(\tau) \left( \fr{1}{3} \pd_\eta^2 \phi 
                        + \fr{4}{3} \pd_\eta \phi \pd_\eta \phi \right) \lap3 \Upsilon_i 
           \nonumber \\
   && + \half M_\P^2 \e^{2\phi} \lap3 \Upsilon_i 
      -\fr{4}{3} \e^{4\phi} \rho \Omega_i =0.
         \label{equation-vector2}
\eea
The second equation is composed of variables with at most the second order of the time derivative, while the first equation is of the third order. Thus, the value of $\Omega_i(\tau_\Lam)$ is determined from the value of $\Upsilon_i(\tau_\Lam)$ as discussed before.

\paragraph{Linear tensor equations}
The linear equation for the gauge invariant tensor perturbation is derived from the space-space component of the stress tensor, $\bT_{ij}=0$, as
\bba
  && -\fr{2}{t_r^2(\tau)} \left\{ \pd_\eta^4 h_{ij}^{\rm TT} -2\lap3 \pd_\eta^2 h_{ij}^{\rm TT} 
                                  + \dlap3 h_{ij}^{\rm TT} \right\} 
             \nonumber \\
  && +\fr{b_1}{8\pi^2}B_0(\tau) \biggl\{
            \left( \fr{1}{3} \pd_\eta^2 \phi + \fr{4}{3}\pd_\eta \phi \pd_\eta \phi  
                   \right) \pd_\eta^2 h_{ij}^{\rm TT}
           + \left( \fr{1}{3} \pd_\eta^3 \phi + \fr{8}{3}\pd_\eta^2 \phi \pd_\eta \phi 
                     \right) \pd_\eta h_{ij}^{\rm TT}
               \nonumber \\
  && \qquad\qquad\qquad
        + \left( -\fr{7}{3}\pd_\eta^2 \phi + \fr{2}{3} \pd_\eta \phi \pd_\eta \phi 
                  \right) \lap3 h_{ij}^{\rm TT} 
            \biggr\}
               \nonumber \\
  && + M_\P^2 \e^{2\phi} \left\{ - \half \pd_\eta^2 h_{ij}^{\rm TT} 
          - \pd_\eta \phi \pd_\eta h_{ij}^{\rm TT} + \half \lap3 h_{ij}^{\rm TT} \right\}=0.
          \label{equation-tensor}
\eea

%%%%%%%%%%%%%%%%%%%%%%%%%%%%%%%%%%%%%%
%%%%   Primordial Power Spectra  %%%%
%%%%%%%%%%%%%%%%%%%%%%%%%%%%%%%%%%%%%%
\section{Primordial Power Spectra}
\setcounter{equation}{0}
\setcounter{figure}{0}
\noindent

The observed CMB anisotropies are understood as a reflection of the Bardeen potentials on the last scattering surface \cite{sw,hs,ll}. Before the universe was neutralized, we can trace the history up to big bang using the Einstein theory. The dynamics we described in the previous section enable us to trace back further to the Planck scale where the inflation ignites.

The low multipole components of the observed CMB anisotropies correspond to the fluctuation of the size of the order of the Hubble distance $1/H_0$, where $H_0$ is the present Hubble constant. If the universe expands more than about the order of $10^{60}$ from the begining during the inflation and the Einstein periods, the fluctuation of the order of the Hubble distance originates from a fluctuation less than the order of the Planck length. The idea of inflation suggests that this order of the expansion is able to solve the flatness and horizon  problems. Thus, we hope to observe the Planck scale physics from the CMB anisotropies.

%%%%%%%%%%%%%(Figure 5.1 input)%%%%%
\begin{figure}[t]
\begin{center}
\includegraphics{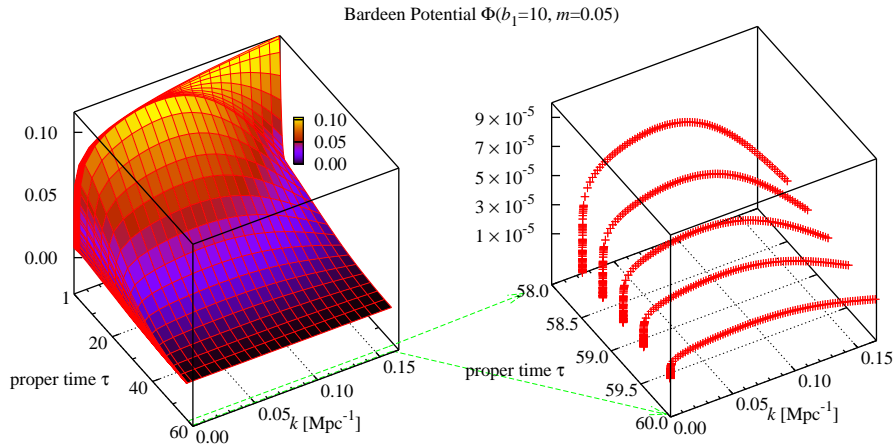}
\end{center}
\caption{\label{fig 5.1} The evolution of Bardeen potential $\Phi$ for $b_1=10$ and $m=0.05$ Mpc$^{-1}$.}
\end{figure}
%%%%%%%%%%%%%%%%%%%%%%%%%%%%%%%%%%%%
%%%%%%%%%%%%%(Figure 5.2 input)%%%%%
\begin{figure}[t]
\begin{center}
\includegraphics{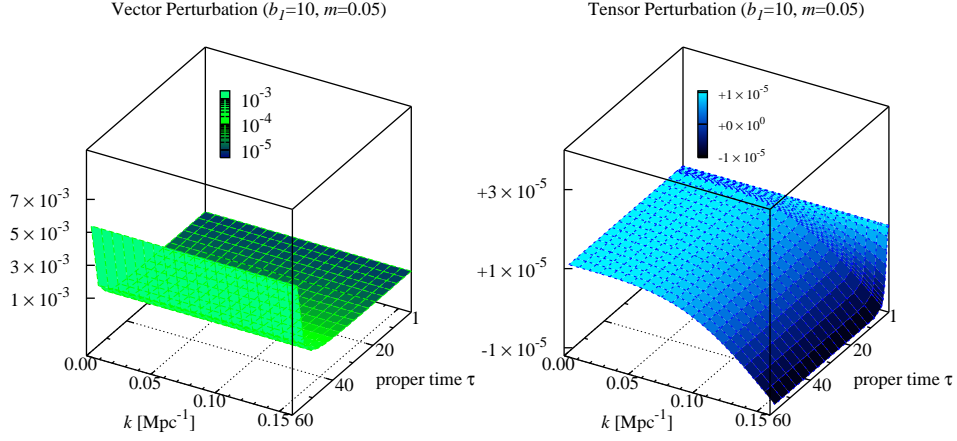}
\end{center}
\caption{\label{fig 5.2} The evolution of vector and tensor perturbations, $\Upsilon_i$ and $h^{\rm TT}_{ij}$, for $b_1=10$ and $m=0.05$ Mpc$^{-1}$. }
\end{figure}
%%%%%%%%%%%%%%%%%%%%%%%%%%%

We solve the coupled system of equations, (\ref{equation-trace}), (\ref{equation-space}), (\ref{equation-vector1}), and (\ref{equation-tensor}), numerically to obtain the values of the Bardeen potentials, the vector and the tensor perturbations. We start the simulation at the Planck time $\tau_\P ~(=1/H_\D)$, where the linear perturbation theory becomes applicable. At this time, the running coupling constant is still small enough so that fluctuations of the conformal field are dominated by the Bardeen potentials satisfying the relation $\Phi =\Psi$, and the vector and the tensor perturbations are relatively small.

The scalar power spectrum is given by the two-point quantum correlation of Bardeen potentials computed by conformal field theory. We write the Fourier transform of a field $f$ in a comoving momentum space as
\bb
     f(\tau, \bx) = \int d^3 \bk f(\tau, \bk) k^{-3} \e^{i\bk \cdot \bx} ,
\ee
and its two-point function as $\langle f(\tau,\bk) f(\tau,\bk^\pp) \rangle = |f(\tau,\bk)|^2 k^3 \dl^3(\bk-\bk^\pp)$. The initial profile of the scalar spectrum is assumed to be \cite{amm-cmb,hy}
\bb
    P^{\rm S}_{\rm pl}(k) = | \Phi(\tau_\P, k) |^2
      = A_{\rm S} \left( \fr{k}{m} \right)^{n_s-1}.
\ee
Here and in the following we write $k=|\bk|$, and $A_{\rm S}$ is a dimensionless constant. $m=a(\tau_\P)H_\D$ is the comoving Planck scale at the Planck time. The spectral index $n_s$ is given by the anomalous dimensions of the scalar curvature,
\bb
      n_s = 5- 8 \fr{1-\sq{1-2/b_1}}{1-\sq{1-4/b_1}} 
        = 1 +2/b_1 +4/b_1^2 + o(1/b_1^3),
\ee
where $b_1$ is given by (\ref{b1}). For the large value of $b_1$, the index approaches to that of the Harrison-Zel'dovich \cite{hz}.

The initial profiles of the vector and the tensor spectra are given by
\bba
      P^{\rm V}_{\rm pl}(k) &=& | \Upsilon(\tau_\P, k) |^2 = A_{\rm V} \left( \fr{k}{m} \right)^{n_v},
                \nonumber \\
      P^{\rm T}_{\rm pl}(k) &=& | h^{\rm TT}(\tau_\P, k) |^2 = A_{\rm T} \left( \fr{k}{m} \right)^{n_t}.
\eea
where $A_{\rm V}$ and $A_{\rm T}$ are small dimensionless constants. The indices are given by
\bb
       n_v = n_t =0,
\ee
because the vector and tensor modes are conformally free fields with logarithmic correlations for the distance in the space coordinate.

The spectrum implies that the fluctuation of the distance $1/m$ at present originates from that of the Planck scale, $1/H_\D$, at the Planck time. Beyond the Planck scale, the non-linear effects of the conformal mode will become significant. In order to include non-linear effects, we have to solve the model exactly or to use a non-perturbative method such as dynamical triangulation \cite{horata-yukawa} for the regime beyond the Planck scale.

%%%%%%%%%%%%%(Figure 5.3 input)%%%%%
\begin{figure}[t]
\begin{center}
\begin{minipage}{6cm}
\includegraphics[width=6cm]{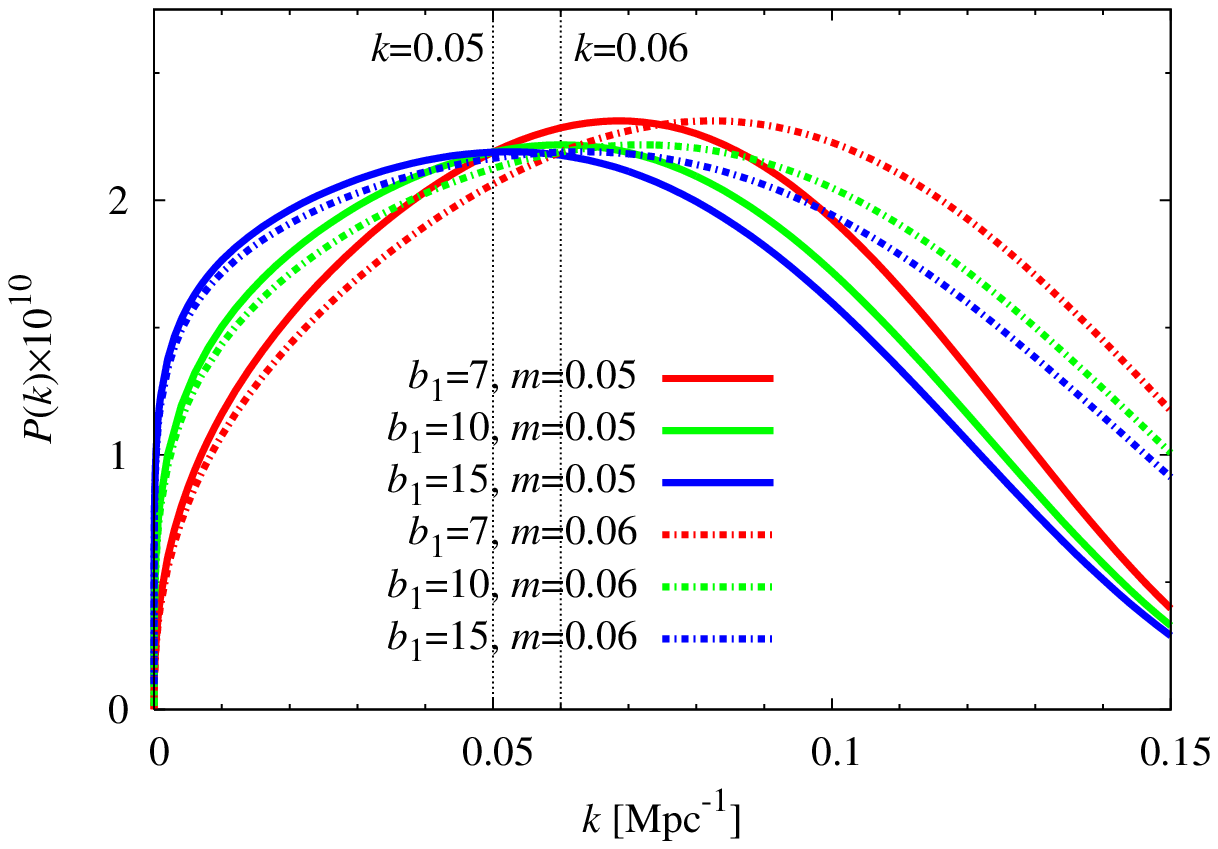}
\end{minipage}
\begin{minipage}{6cm}
 \includegraphics[width=6cm]{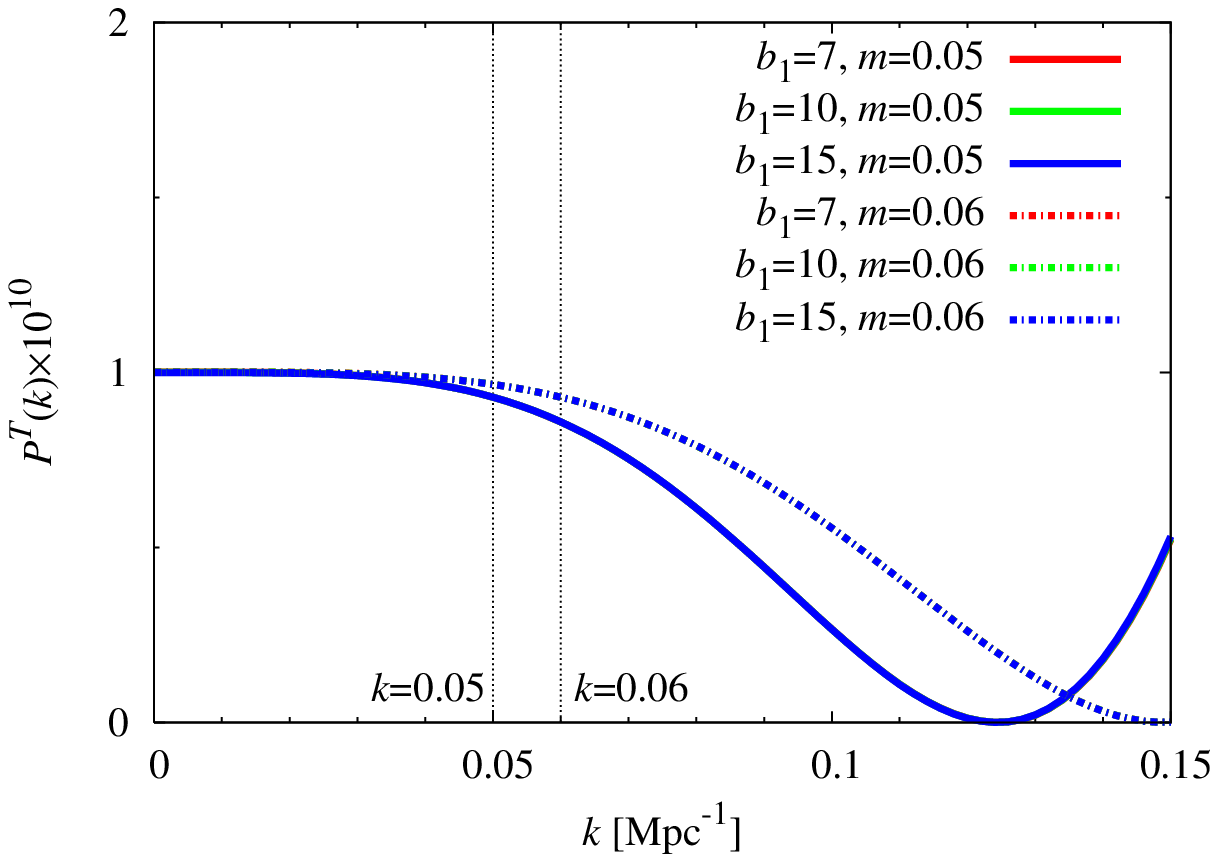}
\end{minipage}
\end{center}
\caption{\label{fig 5.3} The scalar and tensor spectra at the transition point for $b_1=7$, $10$, $15$ with $m=0.05$ and $m=0.06$. The initial amplitudes of $b_1=10$ for each $m$ are set as $\sq{A_{\rm S}}=10^{-1}$ and $\sq{A_{\rm T}}=10^{-5}$. The others are arbitrarily chosen to be equal to the result of $b_1=10$ at $k=m$ for each $m$.}
\end{figure}
%%%%%%%%%%%%%%%%%%%%%%%%%%%%%%%%%
%%%%%%%%%%%%%(Figure 5.4 input)%%%%%
\begin{figure}[t]
\begin{center}
\begin{minipage}{6cm}
\includegraphics[width=6cm]{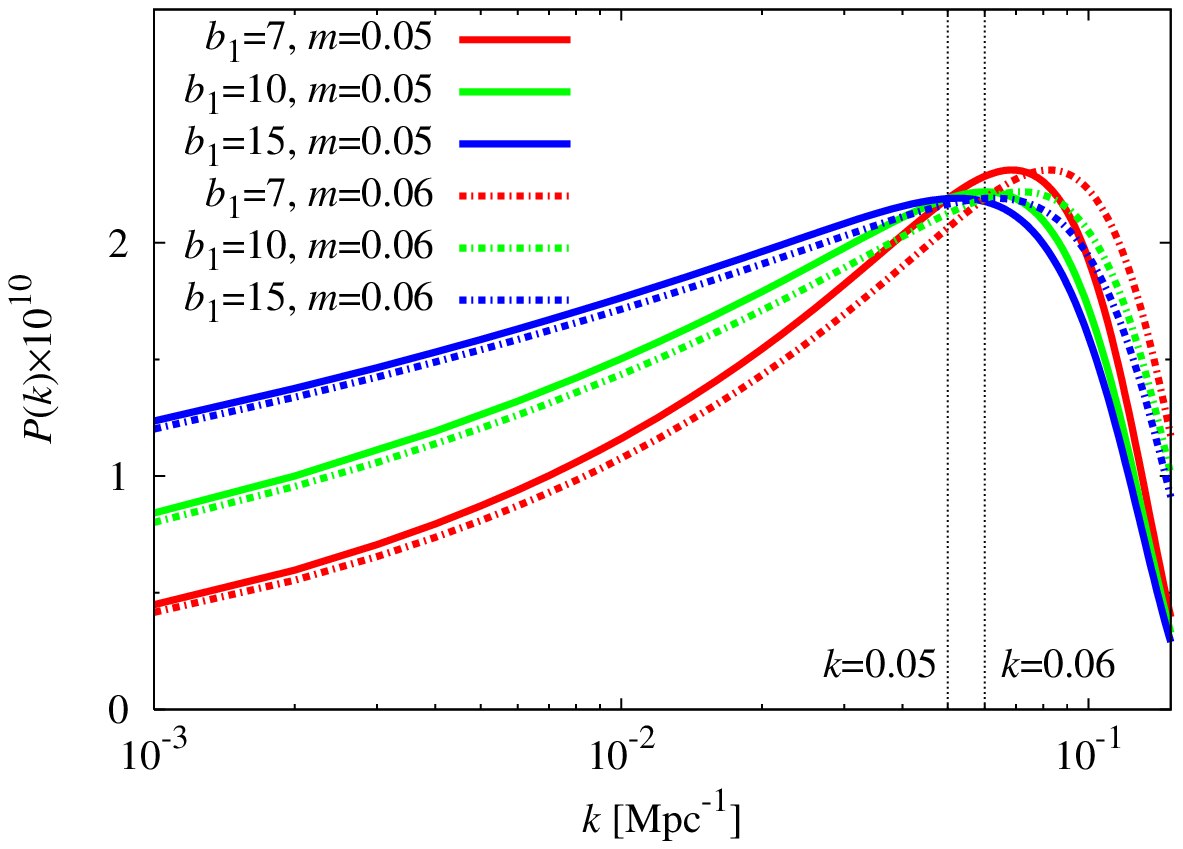}
\end{minipage}
\begin{minipage}{6cm}
 \includegraphics[width=6cm]{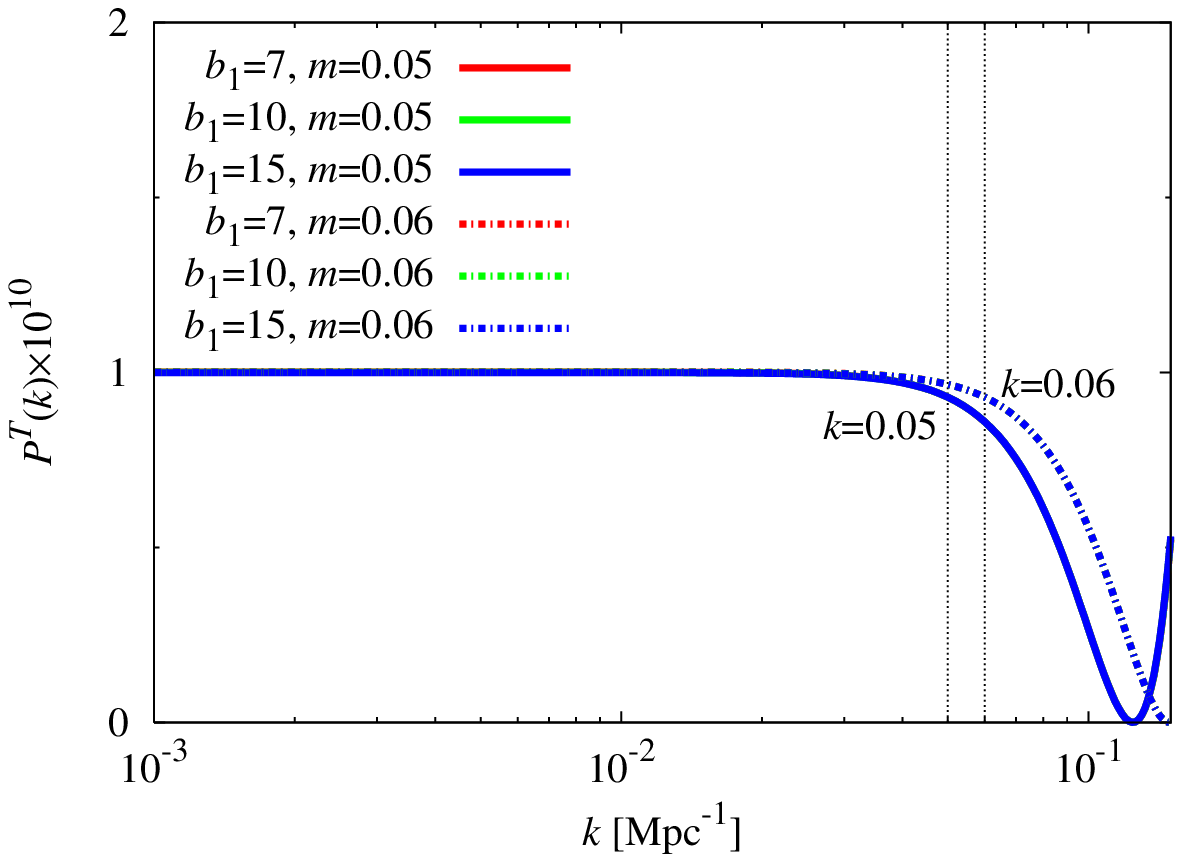}
\end{minipage}
\end{center}
\caption{\label{fig 5.4} The logarithmic plot of the scalar and tensor spectra at $\tau_\Lam$.}
\end{figure}
%%%%%%%%%%%%%%%%%%%%%%%%%%%%%%%%%

The initial values of the Bardeen potentials are set as $\Phi(\tau_\P,k)=\Psi(\tau_\P,k)=\sq{P^{\rm S}_{\rm pl}(k)}$ at the Planck time. Since the Bardeen potentials are fluctuations about the homogeneous solution, the boundary values at $k=0$ satisfy the Dirichlet conditions 
\bb
         \Phi(\tau, 0) = \Psi(\tau, 0) =0.
           \label{dirichlet}
\ee
Since the running coupling diverges at $\tau_\Lam$, equation (\ref{equation-space}) implies that the Bardeen potentials at the transition point satisfy the relation
\bb
       \Phi(\tau_\Lam, k) + \Psi(\tau_\Lam, k)=0.
       \label{boundary}
\ee  
We numerically solve the coupled equations (\ref{equation-trace}) and (\ref{equation-space}) imposing the boundary condtions, (\ref{dirichlet}) and (\ref{boundary}). The initial conditions for the time derivatives of $\Phi$ up to the third order set to zero and $\Psi$ satisfies the boundary (\ref{boundary}) at $\tau_\Lam$. The initial values of the vector and the tensor perturbations are set as $\Upsilon_i (\tau_\P,k)=\sq{A_{\rm V}}$ and $h_{ij}^{\rm TT}(\tau_\P,k) = \sq{A_{\rm T}}$ and the time-derivatives of these fields are also chosen to vanish at $\tau_\P$.

The density and the velocity perturbations are computed using the equations (\ref{equation-energy}) and (\ref{equation-velocity}), respectively. The values at the transition point are calculated using the Bardeen potential to be 
\bba
      D(\tau_\Lam,k)&=& \fr{2}{3}\fr{H_\D^2}{H(\tau_\Lam)^2} 
                           \e^{-2{\cal N}_e} \fr{k^2}{m^2}\Phi(\tau_\Lam, k),
                 \nonumber \\
      k V(\tau_\Lam,k) &=& \half \fr{H_\D}{H(\tau_\Lam)^2} \e^{-{\cal N}_e} \fr{k}{m} 
                      \left( \dot{\Phi}(\tau_\Lam,k) +H(\tau_\Lam)\Phi(\tau_\Lam,k) \right), 
\eea 
where ${\cal N}_e$ is the number of e-foldings defined by (\ref{e-foldings}). Here, we multiply $k$ on $V$ in order to make the expression dimensionless. Also the vector perturbation $\Omega_i$ is computed using equation (\ref{equation-vector2}), resulting
\bb
     \Omega_i(\tau_\Lam,k) = -\fr{1}{8} \fr{H_\D^2}{H(\tau_\Lam)^2} \e^{-2{\cal N}_e} 
                             \fr{k^2}{m^2} \Upsilon_i(\tau_\Lam,k). 
\ee
To derive these values, we use the value $\rho(\tau_\Lam) = 3 M_\P^2 H(\tau_\Lam)^2$. Due to the exponential factor, these values are small enough to solve the flatness problem.

Let us examine the parameters to be used for the simulation. The results of simulation are given in figures \ref{fig 5.1} and \ref{fig 5.2}. The value of the coefficient in front of the Wess-Zumino action is taken to be $b_1=10$, which gives $n_s=1.25$, and the values of other parameters to determine the homogeneous background are the same as those used in section 3. The comoving Planck scale is taken as $m=0.05$ Mpc$^{-1}$. Since $H_\D$ is given by the order of $10^{19}$ GeV, the scale factor at the Planck time $a(\tau_\P)$ becomes the order of $10^{-59}$, where the present scale factor is normalized to be unity. On the other hand, after the transition occured at the energy scale $\Lam_\QG = 1.1 \times 10^{17}$ GeV, where $M_\P=2.4 \times 10^{18}$ GeV, the universe expands about $10^{29}$. This implies that the space-time expands about the order of $10^{30}$ in  the inflationary period starting from the Planck time and a few accelerating expansion period right after the transition before reducing to the Friedmann universe. The number of e-folding for these periods corresponds to $70$. It is consitent with the solution given in section 3. The initial amplitudes of the Bardeen potentials, the vector and the tensor perturbations are taken to be $\sq{A_{\rm S}}=10^{-1}$ and $\sq{A_{\rm V}}=\sq{A_{\rm T}}=10^{-5}$.

The amplitudes of the Bardeen potentials reduces much slower than the damping rate during the inflation. Existence of the gradually varing component can be seen in the analytical solution of approximate equation of motion valid in the intermediate stage of inflation as shown in the appendix C. The tensor fluctuation is preserved to be small, while the vector fluctuation increases near the transition point. However, in the Einstein phase, the scalar and tensor fluctuations survive until present epoch, while the vector fluctuation soon disappear and we cannot observe it. The primordial scalar and tensor spectra evaluated at the transition point are defined by
\bba
     P^{\rm S}(k)&=& |\Phi(\tau_\Lam,k)|^2, 
       \nonumber \\
     P^{\rm T}(k)&=& |h^{\rm TT}(\tau_\Lam,k)|^2.
\eea
The results are given in figures \ref{fig 5.3} and \ref{fig 5.4}, where we plot the cases of $b_1(n_s)=7(1.41)$, $10(1.25)$, $15(1.15)$ with $m=0.05$ and $0.06$. The simulation is carried out for the region $k \leq 0.15$, using the Fortran software, BVP\_SOLVER \cite{bvp-solver}.

It seems that the linear approximation is not applicable for the high momentum region $k^2/m^2 \gg 1$, because such a fluctuation corresponds initially too far beyond the Planck scale so that there appear coefficients with large values in the equations which violate the applicability of the linear approximation. The patterns of the spectra are sensitive to the values of $m$ and $b_1$, while the parameters determing the dynamical factor $B_0$ are insensitive to that, apart from the magnitude of amplitude. The patterns of spectra are also preserved in the Einstein era, although the amplitudes may slightly change in a few moment after the transition.

The initial spectra we employed here do not cut low momentum components required to explain the observed suppression of low l-components in the angular power spectra. In order to explain it, we meight need to seek a reason in the ambiguity of the choice of initial spectra which should reflect original non-perturbative dynamics of the traceless mode for the long-distance two-point correlation functions.

%%%%%%%%%%%%%%%%%%%%%%%%
%%%%   Conclusion   %%%%
%%%%%%%%%%%%%%%%%%%%%%%%
\section{Conclusion}
\setcounter{equation}{0}
\noindent

We have constructed an evolutional model of the universe based on the conformal gravity with the symmetry breaking Einstein action. The universe starts from a quantum state at the Planck time and grows up exponentially. As a consequence of the asymptotically free dynamics, there appears a strong coupling phase where the structure of the space-time changes drastically. The transition takes place at the dynamical energy scale of gravity, where quantum space-time with conformal invariance makes transition to classical Einstein universe. Then the universe grows up to the present size, such that the size of the Planck length at the Planck time extends to the size of the order of $10$ Mpc distance today. We have suggested that at the transition, field fluctuations freeze to localized objects, and they eventually decay into the classical matter driving the universe into the big bang phase.

It was shown that the linear perturbation is applicable on the inflationary background for the momentum range which covers the size of fluctuation observed as CMB anisotropies today by COBE and WMAP. The evolutions of gravitational scalar, vector and tensor fluctuations have been evaluated. Since the initial fluctuations are provided by conformal field theory, we expect that the amplitudes of tensor and vector fluctuations are relatively small in comparison to that of scalar fluctuations. The scalar fluctuations are getting small during the inflation, and the tensor fluctuation is preserved to be small, while the vector fluctuation is getting large near the transition point. However, since the vector fluctuation disappears in the Einstein era, the tensor fluctuation in addition to the scalar fluctuation may contribute to the primordial spectra constituting the observed CMB anisotropies.

The condition $M_\P \gg \Lam_\QG$ is significant to make the inflationary scenario. It also implies that quantum effects turn on the size much larger than the Planck length so that not only the space-time singularity but also the horizon of an elementary excitation with the Planck mass disappear.
Furthermore, the strong repulsive effect in quantum gravity which causes the inflation also erase the singularity inside a black hole by balancing the pressure of the collapsing matter. However, if the collapsing goes too far and exceedes the balance, the black hole may explode something like as mini-inflation.

\vspace{1cm}

%%%%%%%%%%%%%%%%%%
%%%  Appendix  %%%
%%%%%%%%%%%%%%%%%%
\begin{center}
{\Large {\bf Appendix}}
\end{center}

\appendix 

%%%%%%%%%%%%%%%%%%%%%%%%%%%%%%%%%%%%%%%
%%%  Stress Tensor for Each Sector  %%%
%%%%%%%%%%%%%%%%%%%%%%%%%%%%%%%%%%%%%%%
\section{Stress Tensor for Each Sector}
\setcounter{equation}{0}
\noindent

In the appendix A, the conformal field $\phi$ denotes a full field including the homogeneous and non-homogeneous parts, such that $\phi = \phi(\eta,\bx)$ (\ref{non-homo-field}). To obtain the linear perturbation theory, the stress tensor has to be expanded by the fluctuating field $\vphi$.

Equations of motion are obtained by the variations of the effective action as
\bba
  \dl \Gm &=& \half \int d^4 x \sq{-g} T^{\mu\nu} \dl g_{\mu\nu}
             \nonumber \\
          &=& \half \int d^4 x \sq{-\bg} \left\{ 2 \bar{T}^\lam_{~\lam} \dl \phi 
                  +  \bar{T}^{\mu\nu} \dl \bg_{\mu\nu}  \right\} 
             \nonumber \\
    &=& \int d^4 x \left\{ \bT^\lam_{~\lam} \dl\phi 
                          +\half \bT^\mu_{~\nu} \dl h^\nu_{~\mu} \right\} =0,
\eea
where the indices of stress tensor $\bT_{\mu\nu}$ are contracted by the flat background metric $\eta_{\mu\nu}$. The ordinary stress tensor defined by the variation of the physical metric $g_{\mu\nu}$ is denoted by $T_{\mu\nu}$. Furthermore, we write the stress tensor defined by the variation of the metric $\bg_{\mu\nu}$ as $\bar{T}_{\mu\nu}$. Indices of these stress tensors are contracted by $g_{\mu\nu}$ and $\bg_{\mu\nu}$, respectively. The difference of $T_{\mu\nu}$ and $\bar{T}_{\mu\nu}$ is just a conformal factor such that $T^{\mu\nu}=\e^{-6\phi}\bar{T}^{\mu\nu}$ and $T^\mu_{~\nu}=\e^{-4\phi}\bar{T}^\mu_{~\nu}$ and so on. Within linear perturbations in $h^\mu_{~\nu}$, the relation between $\bar{T}_{\mu\nu}$ and $\bT_{\mu\nu}$ is given by the symmetrized product $\bT_{\mu\nu}=\eta_{\lam(\mu}\bar{T}^\lam_{~\nu)}$.

The density and velocity perturbations in the matter sector are defined by the physical stress tensor on the metric $g_{\mu\nu}$ as 
\bba
     T^{{\rm M}0}_{~~~ 0} &=& - (\rho + \dl \rho),
           \nonumber \\
     T^{{\rm M}i}_{~~~ 0} &=& - (\rho + P)v^i,
           \nonumber \\
     T^{{\rm M}0}_{~~~ j} &=& (\rho + P)(v_j+h_{0j}),
           \nonumber \\
     T^{{\rm M}i}_{~~~ j} &=& (P + \dl P) \dl^i_{~j} + \Pi^i_{~j}.
\eea
where $v_i=\hg_{ij}v^j$. For conformal invariant matters, the equation of state is given by $w=P/\rho=1/3$, and $\dl P/\dl \rho =1/3$. The trace of the anisotropic stress vanishes, $\Pi^i_{~i}=0$. The stress tensor is transformed as $\dl_\xi T^{{\rm M}\mu}_{~~~ \nu}=\pd_\nu \xi^\lam T^{{\rm M}\mu}_{~~~ \lam}-\pd_\lam \xi^\mu T^{{\rm M}\lam}_{~~~ \nu}+\xi^\lam \pd_\lam T^{{\rm M}\mu}_{~~~ \nu}$ under the general coordinate transformation.

Hence, the dynamical equations of motion are given by
\bb
    \bT_{\mu\nu} =  \bT^{\rm R}_{\mu\nu} + \bT^{\rm W}_{\mu\nu} 
                     +\bT^{\rm EH}_{\mu\nu} + \bT^{\rm M}_{\mu\nu} = 0,
\ee
where ${\rm R}$, ${\rm W}$ and ${\rm EH}$ denote the stress tensor for the Riegert action, the Weyl action and the Einstein action, respectively. They are given below, while $\bT^{\rm M}_{\mu\nu}$ is given by (\ref{T-matter}).

\paragraph{Riegert action}
\bba
 \bT^{\rm R}_{\mu\nu}
  &=& \fr{b_1}{8\pi^2} \Biggl\{ 
  4\Box \phi \pd_\mu\pd_\nu \phi 
  -4 \pd_{(\mu}\Box\phi \pd_{\nu)}\phi 
  -\fr{8}{3}\pd_\mu\pd_\lam \phi \pd_\nu\pd^\lam \phi 
  +\fr{4}{3}\pd_\mu \pd_\nu \pd_\lam \phi \pd^\lam \phi
 \nonumber \\ && 
  +\fr{2}{3}\pd_\mu \pd_\nu \Box \phi 
  +\eta_{\mu\nu} \left[ -\Box\phi \Box\phi 
  +\fr{2}{3} \pd_\lam \Box \phi \pd^\lam \phi 
  +\fr{2}{3} \pd_\lam \pd_\s \phi \pd^\lam \pd^\s \phi 
  -\fr{2}{3} \Box^2 \phi \right] 
 \nonumber \\ && 
  -4 \pd_{(\mu}h^\lam_{\nu)} \Box\phi \pd_\lam \phi 
  +2 \pd^\lam h_{\mu\nu} \Box\phi \pd_\lam \phi 
  -4 h^{\lam\s} \pd_\lam \pd_\s \phi \pd_\mu\pd_\nu \phi 
  -4 \chi^\lam \pd_\mu \pd_\nu \phi \pd_\lam \phi
 \nonumber \\ && 
  +4 h^{\lam\s} \pd_\lam \pd_\s \pd_{(\mu}\phi \pd_{\nu)}\phi 
  +4 \pd_{(\mu}h^{\lam\s} \pd_\lam \pd_\s \phi \pd_{\nu)} \phi 
  +4\chi^\lam \pd_\lam \pd_{(\mu}\phi \pd_{\nu)}\phi 
 \nonumber \\ && 
  +\fr{8}{3} h^{\lam\s} \pd_\mu \pd_\lam \phi \pd_\nu \pd_\s \phi  
  +\fr{4}{3} \pd_{(\mu} h^{\lam\s} \pd_{\nu)} \pd_\lam \phi \pd_\s \phi 
  +4 \pd^\lam h^\s_{(\mu} \pd_{\nu)}\pd_\lam \phi \pd_\s \phi
 \nonumber \\ && 
  -4 \pd^\lam h^\s_{(\mu} \pd_{\nu)} \pd_\s \phi \pd_\lam \phi 
  -\fr{4}{3} h^{\lam\s} \pd_\mu \pd_\nu \pd_\lam \phi \pd_\s \phi 
  -\fr{4}{3} \pd_{(\mu}h^\lam_{\nu)} \pd_\lam \pd_\s \phi \pd^\s \phi
 \nonumber \\ &&
  +\fr{2}{3} \pd^\lam h_{\mu\nu} \pd_\lam \pd_\s \phi \pd^\s \phi 
  +\fr{4}{3} \pd_\mu \pd_\nu h^{\lam\s} \pd_\lam \phi \pd_\s \phi 
  -4 \pd^\lam \pd_{(\mu} h^\s_{\nu)} \pd_\lam \phi \pd_\s \phi
 \nonumber \\ &&
  +2 \pd^\lam \pd^\s h_{\mu\nu} \pd_\lam \phi \pd_\s \phi
  -4 \pd^\lam \chi_{(\mu} \pd_{\nu)} \phi \pd_\lam \phi
  +4 \Box h^\lam_{(\mu} \pd_{\nu)} \phi \pd_\lam \phi
 \nonumber \\ &&
  +\fr{4}{3} \pd_{(\mu} \chi_{\nu)} \pd_\lam \phi \pd^\lam \phi
  -\fr{2}{3} \Box h_{\mu\nu} \pd_\lam \phi \pd^\lam \phi 
  +\fr{4}{3} \pd_\lam \chi^\lam \pd_\mu \phi \pd_\nu \phi
 \nonumber \\ &&
  -4 h^\lam_{(\mu} \pd_{\nu)} \pd_\lam \phi \Box \phi
  +2 h^\lam_{(\mu} \pd_{\nu)} \Box \phi \pd_\lam \phi
  +2 h_{\lam (\mu} \pd^\lam \Box \phi \pd_{\nu)} \phi
 \nonumber \\ &&
  +\fr{8}{3} h^\lam_{(\mu} \pd_{\nu)} \pd^\s \phi \pd_\lam \pd_\s \phi  
  -\fr{4}{3} h^\lam_{(\mu} \pd_{\nu)} \pd_\lam \pd_\s \phi \pd^\s \phi
 \nonumber \\ &&
  + \eta_{\mu\nu} \biggl[ 
     2h^{\lam\s} \pd_\lam \pd_\s \phi \Box \phi 
     +2\chi^\lam \Box \phi \pd_\lam \phi 
     -\fr{2}{3}h^{\lam\s} \pd_\lam \pd_\s \pd_\rho \phi \pd^\rho \phi 
 \nonumber \\ && \quad
     -\fr{2}{3}\chi^\lam \pd_\lam \pd_\s \phi \pd^\s \phi 
     + 2 \pd^\lam \chi^\s \pd_\lam \phi \pd_\s \phi 
     -\fr{2}{3} h^{\lam\s} \pd_\lam \Box \phi \pd_\s \phi
     -\fr{4}{3} h^{\lam\s} \pd_\lam \pd_\rho \phi \pd_\s \pd^\rho \phi
 \nonumber \\ && \qquad\quad
     -\fr{4}{3} \pd^\rho h^{\lam\s} \pd_\rho \pd_\lam \phi \pd_\s \phi 
     -\fr{4}{3} \Box h^{\lam\s} \pd_\lam \phi \pd_\s \phi 
     -\fr{2}{3} \pd_\lam \chi^\lam \pd_\s \phi \pd^\s \phi  
     \biggr]
 \nonumber \\ &&
  -\fr{4}{3} \pd_{(\mu} h^{\lam\s} \pd_{\nu)} \pd_\lam \pd_\s \phi 
  -\fr{2}{3} h^{\lam\s} \pd_\lam \pd_\s \pd_\mu \pd_\nu \phi
  -\fr{8}{3} \pd_\mu \pd_\nu h^{\lam\s} \pd_\lam \pd_\s \phi
  -\fr{2}{3} \chi^\lam \pd_\lam \pd_\mu \pd_\nu \phi
 \nonumber \\ &&
  +\fr{8}{3} \pd_{(\mu} \chi^\lam \pd_{\nu)} \pd_\lam \phi 
  -\fr{2}{3} \pd_\mu \pd_\nu \chi^\lam \pd_\lam \phi
  -\fr{2}{3} \pd_{(\mu} h^\lam_{\nu)} \pd_\lam \Box \phi 
  +\fr{1}{3} \pd^\lam h_{\mu\nu} \pd_\lam \Box \phi
 \nonumber \\ &&
  +4 \pd^\lam \pd_{(\mu} h^\s_{\nu)} \pd_\lam \pd_\s \phi
  -2 \pd^\lam \pd^\s h_{\mu\nu} \pd_\lam \pd_\s \phi
  -\fr{14}{3} \pd_{(\mu} \chi_{\nu)} \Box \phi
  +\fr{7}{3} \Box h_{\mu\nu} \Box \phi
 \nonumber \\ &&
  -2 \pd_\lam \chi^\lam \pd_\mu \pd_\nu \phi
  +4 \pd_\lam \chi_{(\mu} \pd_{\nu)} \pd^\lam \phi
  -4 \Box h^\lam_{(\mu} \pd_{\nu)} \pd_\lam \phi
  +\fr{2}{3} \pd_\lam \pd_{(\mu} \chi^\lam \pd_{\nu)} \phi
 \nonumber \\ &&
  -\fr{2}{3} h^\lam_{(\mu} \pd_{\nu)} \pd_\lam \Box \phi
 \nonumber \\ &&
  + \eta_{\mu\nu} \biggl[ 
       \fr{4}{3} h^{\lam\s} \pd_\lam \pd_\s \Box \phi 
       +\fr{4}{3} \pd^\rho h^{\lam\s} \pd_\rho \pd_\lam \pd_\s \phi
       +\fr{4}{3} \chi^\lam \pd_\lam \Box \phi
       +\fr{8}{3} \Box h^{\lam\s} \pd_\lam \pd_\s \phi
 \nonumber \\ && \qquad\quad
      -\fr{8}{3} \pd^\lam \chi^\s \pd_\lam \pd_\s \phi 
      +\fr{2}{3} \Box \chi^\lam \pd_\lam \phi
      +2 \pd_\lam \chi^\lam \Box \phi
      -\fr{1}{3} \pd^\lam \pd_\s \chi^\s \pd_\lam \phi
       \biggr] 
 \nonumber \\ &&
  -\fr{1}{9} \pd_\mu \pd_\nu \pd_\lam \chi^\lam
  +\fr{1}{9} \eta_{\mu\nu} \Box \pd_\lam \chi^\lam   
      \Biggr\},
\eea
where $\chi_\mu = \pd_\lam h^\lam_{~\mu}$ and $\Box=\pd^\lam \pd_\lam = -\pd_\eta^2+\lap3$. The trace of this tensor is given by
\bba
 \bT^{{\rm R} \lam}_{~~~\lam}
 &=& \fr{b_1}{8\pi^2} \left( -2\bDelta_4 \phi -\half \bE_4 \right) 
         \nonumber \\
 &=& \fr{b_1}{8\pi^2} \biggl\{ 
      -2 \Box^2 \phi 
      +4 h^{\mu\nu} \Box \pd_\mu \pd_\nu \phi 
      +4 \pd^\lam h^{\mu\nu} \pd_\lam \pd_\mu \pd_\nu \phi 
      +4 \chi^\lam \pd_\lam \Box \phi 
 \nonumber \\ && \quad
      +4 \Box h^{\mu\nu} \pd_\mu \pd_\nu \phi
      +2 \Box \chi^\lam \pd_\lam \phi
      +\fr{4}{3} \pd_\lam \chi^\lam \Box \phi
      -\fr{2}{3} \pd^\lam \pd_\s \chi^\s \pd_\lam \phi
 \nonumber \\ && \quad
     +\fr{1}{3} \Box \pd_\lam \chi^\lam 
      \biggr\}.
\eea
It is also derived by the variation of the conformal mode.

\paragraph{Weyl action}
\bba
\bT^{\rm W}_{\mu\nu}&=& \fr{1}{t^2} 8 \pd^\lam \pd^\s C^{linear}_{\mu\lam\nu\s} 
     \nonumber \\
   &=& 
          -\fr{2}{t^2} \left\{ 
               \Box^2 h_{\mu\nu} -2 \Box \pd_{(\mu} \chi_{\nu)} 
               +\fr{2}{3} \pd_\mu \pd_\nu \pd_\lam \chi^\lam 
               +\fr{1}{3} \eta_{\mu\nu}\Box \pd_\lam \chi^\lam 
               \right\}.
\eea
This tensor is traceless: $\bT^{{\rm W} \lam}_{~~~\lam}=0$.

\paragraph{Einstein action}
\bba
 \bT^{\rm EH}_{\mu\nu} 
  &=& M_{\rm P}^2 \e^{2\phi} \biggl\{ 
    2 \pd_\mu \pd_\nu \phi                    -2 \pd_\mu \phi \pd_\nu \phi
    +\eta_{\mu\nu} \left( -2 \Box \phi -\pd^\lam \phi \pd_\lam \phi \right)
 \nonumber \\ &&
  -\pd_{(\mu} \chi_{\nu)}                     +\half \Box h_{\mu\nu}
  -2 h^\lam_{(\mu} \pd_{\nu)} \pd_\lam \phi   +2 h^\lam_{(\mu} \pd_{\nu)} \phi \pd_\lam \phi
 \nonumber \\ &&
  -2 \pd_{(\mu} h^\lam_{\nu)} \pd_\lam \phi   +\pd^\lam h_{\mu\nu} \pd_\lam \phi
 \nonumber \\ &&
  +\eta_{\mu\nu} \left( \half \pd_\lam \chi^\lam               +2 h^{\lam\s} \pd_\lam \pd_\s \phi
                        +h^{\lam\s} \pd_\lam \phi \pd_\s \phi   +2 \chi^\lam \pd_\lam \phi     \right)
  \biggr\}.
\eea
The trace of the stress tensor is given by
\bba
 \bT^{{\rm EH} \lam}_{~~~~\lam}
  &=& M_{\rm P}^2 \e^{4\phi} R 
   = M_{\rm P}^2 \e^{2\phi} \left( 
           {\bar R} -6 \overline{\nabla}^2 \phi -6 \overline{\nabla}^\lam \phi \overline{\nabla}_\lam \phi 
           \right)
                   \nonumber \\ 
  &=& M_{\rm P}^2 \e^{2\phi} \Bigl\{
   -6 \Box \phi                         -6 \pd^\lam \phi \pd_\lam \phi
   +\pd_\lam \chi^\lam                  +6 h^{\lam\s} \pd_\lam \pd_\s \phi
                   \nonumber \\
  && \qquad\qquad
   +6 \chi^\lam \pd_\lam \phi           +6 h^{\lam\s} \pd_\lam \phi \pd_\s \phi
         \Bigr\}.
\eea

%%%%%%%%%%%%%%%%%%%%%%%%%%%%%%%%%%%%%%
%%%    Proper Time Representation  %%%
%%%%%%%%%%%%%%%%%%%%%%%%%%%%%%%%%%%%%%
\section{Proper Time Representation}
\setcounter{equation}{0}
\noindent

The proper time is defined by the equation $d\tau =a(\tau) d\eta$, where $a(\tau)=\e^{\phi(\tau)}$ is a solution of the homogeneous equation of motion. Then, we obtain
\bba
  \lap3 &=& a^2 \left( - \fr{k^2}{a^2} \right), 
        \nonumber \\
  \pd_\eta &=& a \pd_\tau, 
        \nonumber \\
  \pd_\eta^2 &=& a^2 \left( \pd_\tau^2 + H \pd_\tau \right), 
        \nonumber \\
  \pd_\eta^3 &=& a^3 \left\{ \pd_\tau^3 +3 H \pd_\tau^2 + \left( \dH +2 H^2 \right) \pd_\tau \right\}, 
        \nonumber \\
  \pd_\eta^4 &=& a^4 \left\{ \pd_\tau^4 +6 H \pd_\tau^3 + \left( 4 \dH +11 H^2 \right) \pd_\tau^2 
                             +\left( \ddH + 7H \dH + 6 H^3 \right) \pd_\tau    \right\}  
        \nonumber \\ &&
\eea
and
\bba
  \pd_\eta \phi &=& a H, 
      \nonumber \\
  \pd_\eta^2 \phi &=& a^2 \left( \dH + H^2 \right), 
      \nonumber \\
  \pd_\eta^3 \phi &=& a^3 \left( \ddH  +4 H \dH + 2H^3 \right), 
      \nonumber \\
  \pd_\eta^4 \phi &=& a^4 \left( \dddH +7 H \ddH +4 \dH^2 +18 H^2 \dH +6 H^4 \right) ,
\eea
where $H(\tau)=\dot{a}(\tau)/a(\tau)$.

%%%%%%%%%%%%%%%%%%%%%%%%%%%%%%%%%%%%%%%%%%%%%%%%%%%%%%
%%%%   Analytic Study of Linear Scalar Equations   %%%
%%%%%%%%%%%%%%%%%%%%%%%%%%%%%%%%%%%%%%%%%%%%%%%%%%%%%%
\section{Analytical Study of Linear Scalar Equations}
\setcounter{equation}{0}
\noindent

In this appendix, we study solutions of the coupled linear scalar equations, (\ref{equation-trace}) and (\ref{equation-space}), analytically. We here simplify the equations as follows: the coupling $t_r$ is a small constant such that the Hubble parameter are constant normalized to be $H=H_\D/\sq{B_0}=1$ and $T=b_1B_0 t_r^2/8\pi^2 \ll 1$. Furthermore, we neglect the momentum dependence. Such a situation will be realized in several times after the inflation starts because the momentum dependence, $k/a$, becomes negligible as the scale factor grows. Then, we obtain the following equations for the scalar perturbations:
\bba
  && -2\ddddPhi -14 \dddPhi -36 \ddPhi -48 \dPhi 
     + 2 \dddPsi +14 \ddPsi + 36 \dPsi +48 \Psi
            \nonumber \\ 
  && \qquad 
     +  6 \left( \ddPhi +4 \dPhi - \dPsi -4 \Psi \right) =0,
          \label{trace-simple} \\
  && \fr{4}{3} \ddPhi + \fr{16}{3} \dPhi +\fr{20}{3} \Phi 
     -\fr{4}{3} \dPsi + \fr{4}{3} \Psi 
            \nonumber \\
  && \qquad  
    +\fr{8}{T} \left( \ddPhi +  \dPhi - \ddPsi - \dPsi \right)
     - 2 ( \Phi + \Psi ) =0.
     \label{space-simple}
\eea
Introducing the variable $f=\Psi- \dPhi$, these equations are written as
\bba
   \dddf +7 \ddf +15 \df +12 f &=& 0,
                  \\
   \dddPhi -\left( 1+\fr{7}{12}T \right) \dPhi -\fr{7}{12}T \Phi 
      &=& - \ddf - \left( 1 +\fr{1}{6}T \right) \df -\fr{1}{12}T f.
\eea
The first equation has the solution
\bb
    f = c_1 \e^{-4\tau} +c_2 \e^{-\fr{3}{2}\tau} \sin \left( \fr{\sq{3}}{2}\tau \right)
           +c_3 \e^{-\fr{3}{2}\tau} \cos \left( \fr{\sq{3}}{2}\tau \right).
\ee
Substituting this solution into the second equation, we obtain
\bba
   \Phi &=& (a_1+c_1)\e^{-\tau} +(a_2+c_2)\left( 1-\fr{7}{12}T \tau \right)
          + (a_3+c_3) \left(1+\fr{7}{12}T \tau \right) \e^\tau
             \nonumber \\
        && + c_1 \fr{360-7T}{1800} \e^{-4\tau}
         +\fr{\sq{3}c_2 +5c_3}{14} \e^{-\fr{3}{2}\tau}\cos \left( \fr{\sq{3}}{2}\tau \right) 
              \nonumber \\
        &&  +\fr{5c_2 -\sq{3}c_3}{14} \e^{-\fr{3}{2}\tau}\sin \left( \fr{\sq{3}}{2}\tau \right).
               \label{solution-simple}
\eea
Here, we expand the time constant in $T$ and retain within the first order.

Since we now consider the exactly zero-momentum case, this solution includes both the homogeneous vacuum mode and the fluctuation mode we seek. At $T=0$, the vacuum mode $\Phi=\Psi=\om$ satisfies the following equation:
\bb
     \ddddom+ 6 \dddom +8 \ddom -3 \dom -12 \om =0,
\ee
which is obtained from equation (\ref{trace-simple}), while the second equation (\ref{space-simple}) is trivial in this case. This equation has  an inflationary mode, $\e^\tau$, and three decaying modes, $\e^{-4\tau}$, $\e^{-3\tau/2}\sin(\sq{3}\tau/2)$ and $\e^{-3\tau/2}\cos(\sq{3}\tau/2)$. Thus, by definition, the homogeneous vacuum modes, which behave like these at the $T=0$ limit, must be removed from solution (\ref{solution-simple}). Thus, the fluctuation, $\Phi$, behaves for $T \ll 1$ as 
\bb
    \Phi \sim 1-\fr{7}{12}T\tau,
\ee
where the exponential damping mode is neglected. This mode appears at the intermediate region in time evolution where the momentum dependence becomes negligible (see figure \ref{fig 5.1}).

%%%%%%%%%%%%%%%%%%%%%%
%%%   References   %%%
%%%%%%%%%%%%%%%%%%%%%%

\end{document}